\documentclass[11pt]{article}   
\usepackage{hyperref}
\pdfoutput=1
\usepackage{amsmath}
\usepackage{amssymb}
\usepackage{graphicx}
\usepackage{color}

\begin{document}


\title{Smoothed Boundary Method for Solving Partial Differential Equations with General Boundary Conditions on Complex Boundaries}
\author{Hui-Chia Yu, Hsun-Yi Chen, and K. Thornton \\ Department of Materials Science and Engineering, \\ University of Michigan, Ann Arbor, MI 48109, USA}
\date{Oct 8, 2009}

\maketitle

\begin{abstract}
In this article, we describe an approach for solving partial differential equations with general boundary conditions imposed on arbitrarily shaped boundaries.  A function that has a prescribed value on the domain in which a differential equation is valid and smoothly but rapidly varying values on the boundary where boundary conditions are imposed is used to modify the original differential equations.  The mathematical derivations are straight forward, and generically applicable to a wide variety of partial differential equations.  To demonstrate the general applicability of the approach, we provide four examples: (1) the diffusion equation with both Neumann and Dirichlet boundary conditions, (2) the diffusion equation with surface diffusion, (3) the mechanical equilibrium equation, and (4) the equation for phase transformation with additional boundaries.  The solutions for a few of these cases are validated against corresponding analytical and semi-analytical solutions.  The potential of the approach is demonstrated with five applications: surface-reaction diffusion kinetics with a complex geometry, Kirkendall-effect-induced deformation, thermal stress in a complex geometry, phase transformations affected by substrate surfaces, and a self-propelling droplet.
\end{abstract}


\section{Introduction}

The smoothed boundary method \cite{Bueno-Orovio:2006a,Bueno-Orovio:2006b,Bueno-Orovio:2006c} and other similar approaches \cite{Gal:2006,Wu:2007,Gal:2008} have recently been demonstrated as powerful tools for solving various partial differential equations with boundary conditions imposed within the computational domain. The method's origin can be traced to the embedded boundary method and the immersed boundary method (for an overview, see Ref. \cite{Badea:2001,Peskin:2002,Boyd:2005,Lui:2009,Sabetghadam:2009}).  This method has been successfully employed in simulating diffusion processes \cite{Kockelkoren:2003,Levine:2005} and wave propagation \cite{Bueno-Orovio:2006a,Bueno-Orovio:2006b,Bueno-Orovio:2006c,Fenton:2005,Buzzard:2007} constrained within geometries described by a continuously transitioning domain indicator function (hereafter, the domain parameter) with a no-flux boundary condition imposed on the diffuse interface (as defined by the narrow transitioning region of the domain parameter). While those works demonstrated the potential for this type of numerical methods that circumvents the difficulties with constructing the finite element mesh (e.g., meshing the surface and then building a volumetric mesh based on the surface mesh or by combining regular subdomains that can be easily meshed), which is particularly useful when dealing with complex structures.  However, the method was only applicable to no-flux boundary conditions, and no approaches to extend the method to other types of equations or boundary conditions were available.  Recently, a different formulation, based on asymptotic analyses, to solve partial differential equations in a similar manner was proposed \cite{Li:2009,Lowengrub:2009,Ratz:2006,Teigen:2009a,Sohn:2009,Teigen:2009b}, providing a justification of the method as well as increasing the applicability of the approach.  

In this paper, we provide a mathematically consistent smoothed boundary method and provide a precise derivation for the equations.  The specific equations that we consider are: (1) the diffusion equation with Neumann and/or Dirichlet boundary conditions, (2) the bulk diffusion equation coupled with surface diffusion, (3) the mechanical equilibrium equation for linear elasticity, and (4) Allen-Cahn or Cahn-Hilliard equations with contact angles as boundary conditions.  The method is especially useful for three-dimensional image-based simulations.

\section{Background}

The method is based on a diffuse interface description of different phases, similar to the continuously transitioning order parameters in the phase-field method \cite{Cahn:1958,Cahn:1959,Allen:1979,Ginzburg:1950,Chen:2002,Emmerich:2003} often used in studying phase transformations and microstructural evolution in materials. In phase-field models, phases (which could be liquid, solid, vapor, or two different solids/liquids having different compositions) are described by one or more order parameters having a prescribed bulk values within each phase.  In the interface, the order parameter changes in a controlled manner.  Asymptotic analyses \cite{Emmerich:2003} can be used to show that the phase-field governing equations approach the corresponding sharp interface problems in the sharp interface limit.

\begin{figure}[htb] 
\begin{center}
\includegraphics[width=1\textwidth]{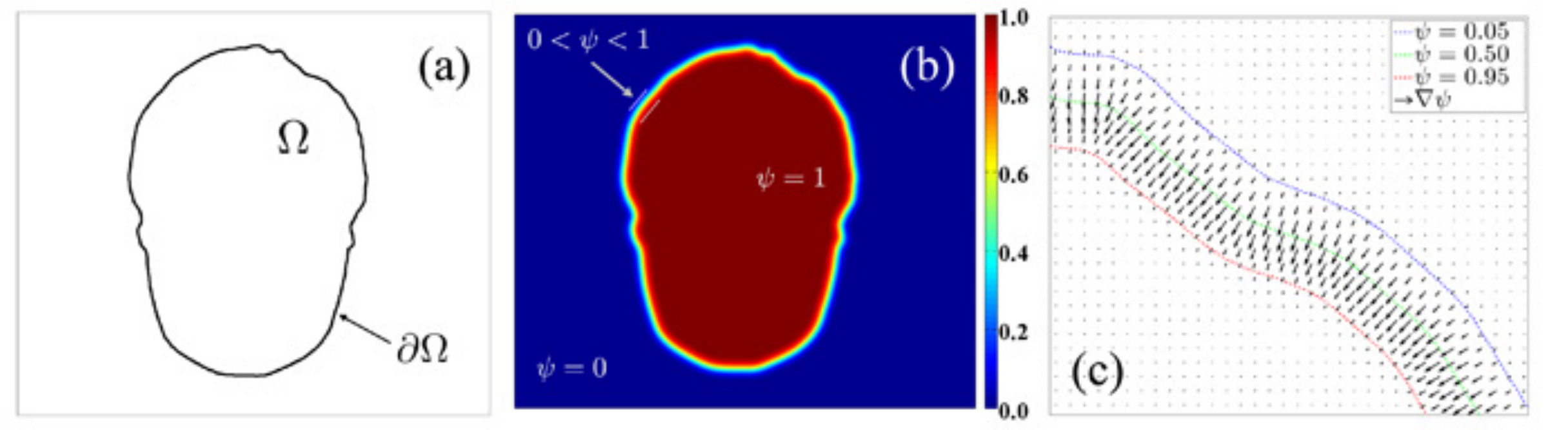}
\end{center}
\caption{(a) The conventional sharp interface description of a domain with a zero-thickness boundary. (b) The diffuse interface domain and boundary defined by a domain parameter, $\psi$. (c) The inward normal vectors defined by $\nabla \psi$.}
\label{Domain}
\end{figure}

We adopt this concept to describe internal domain boundaries by an order-parameter-like domain parameter, which may or may not be stationary and takes a value of 1 inside the domain of interest and 0 outside.  The equations will be solved where the domain parameter is 1, with boundary conditions imposed where the domain parameter is at the intermediate value (approximately 0.5). Figure~\ref{Domain} illustrates a schematic diagram of the sharp and diffuse interfaces. In the conventional sharp interface description, the domain of interest is $\Omega$ and is bound by a zero-thickness boundary denoted by $\partial \Omega$ [Fig.~\ref{Domain}(a)]. Within $\Omega$, the partial differential equations need to be solved according to the boundary conditions imposed at $\partial \Omega$. In the diffuse interface description, we employ a continuous domain parameter, which is uniformly 1 within the domain of interest and uniformly 0 outside. In this case, the originally sharp domain boundary is smeared to a diffuse interface with a finite thickness indicated by $0<\psi<1$. Our target is to solve partial differential equations within the region where $\psi=1$ while imposing boundary conditions at the narrow transitioning interface region where $0<\psi<1$. By using this description, there is no specifically defined domain boundary. The system will determine the boundary by a variation of the domain parameter. In addition, the gradient of the domain parameter $\nabla \psi$ will automatically determine the inward normal vector of the contour level sets of $\psi$ (see Fig.~\ref{Domain}(c)).

\section{Formulation}

\subsection{General Approach}

The general approach is as follows. The domain parameter describes the domain of interest ($\psi=1$ inside the domain, and $\psi=0$ outside).  The transition between the two values described is smooth and taken as the solution to an Allen-Cahn type dynamic equation (having a form of a hyperbolic function) described later. To derive the smoothed boundary formulation for Neumann boundary condition, the differential equation of interest ($H$) is multiplied by the domain parameter, $\psi$.  By using identities of the product rule of differentiation such as
\begin{equation} \label{ChainRule1}
\psi \nabla^2 H =  \nabla \cdot (\psi \nabla H) - \nabla \psi \cdot \nabla H,
\end{equation}
we obtain terms proportional to $\nabla \psi$. Since the unit (inward) normal of the boundary, $\vec{n}$, is given by $\nabla \psi/ |\nabla \psi |$, such terms can be written in $\partial H/\partial n = \nabla H \cdot \vec{n}  = \nabla H \cdot \nabla \psi /|\nabla \psi|$, and thus reformulated to be the Neumann boundary condition imposed on the diffuse interface.  

Similarly, to derive the smoothed boundary formulation for the Dirichlet boundary condition, the equation of interest is multiplied by the square of the domain parameter. Again using mathematical identities, $\psi^2 \nabla^2 H = \psi \nabla \cdot (\psi \nabla H) - \psi \nabla \psi \cdot \nabla H$ where $\psi \nabla \psi \cdot \nabla H =  \nabla \psi \cdot \nabla \left( \psi H \right) - H \left| \nabla \psi \right|^{2}$, we obtain 
\begin{equation} \label{ChainRule2}
\psi^2 \nabla^2 H = \psi \nabla \cdot (\psi \nabla H) - [\nabla \psi \cdot \nabla \left( \psi H \right)-H |\nabla \psi|^2].
\end{equation}
Note that $H = H|_{\partial \Omega}$ associated with $|\nabla \psi|^2$ appearing in the last term is the boundary value imposed on the diffuse interface.

Specific details of the derivation depend on the equation to which the approach is applied, and we therefore provide four examples below.

\subsection{Diffusion Equation} \label{DiffEqn}

The first example is the diffusion equation with Neumann and/or Dirichlet boundary conditions. The Neumann boundary condition is appropriate, for example, as the no-flux boundary condition, while the Dirichlet boundary condition is necessary when the diffusion equation is solved with a fixed concentration on the boundaries.  For Fick's Second Law of diffusion, the original governing equation is expressed as 
\begin{equation}
\frac{\partial C}{\partial t} = -\nabla \cdot \vec{j} + S = \nabla \cdot (D \nabla C) + S, \label{OGE2}
\end{equation}
where $\vec{j}$ is the flux vector, $D$ is the diffusion coefficient, $C$ is the concentration, $S$ is the source term, and $t$ is time. Instead of directly solving the diffusion equation, we multiply both sides of Eq.~\eqref{OGE2} by the domain parameter $\psi$ that describes the domain of the solid phase:   
\begin{equation}
\psi \frac{\partial C}{\partial t} = \psi \nabla \cdot (D \nabla C) + \psi S. \label{MGE5}
\end{equation}
Using the identity $\psi \nabla \cdot (D \nabla C) = \nabla \cdot (\psi D \nabla C) - \nabla \psi \cdot (D \nabla C)$, Eq.\ (\ref{MGE5}) becomes
\begin{equation}
\psi \frac{\partial C}{\partial t} = \nabla \cdot (\psi D \nabla C) - \nabla \psi \cdot (D \nabla C) + \psi S. \label{MGE6}
\end{equation}
Now, let us consider the boundary condition in this formulation. The Neumann boundary condition is the inward flux across the domain boundary, mathematically the normal gradient of $C$ at the diffuse interface, and is treated as
\begin{equation} \label{NBC1}
\vec{n} \cdot \vec{j} = \frac{\nabla \psi}{\left| \nabla \psi \right|} \cdot \vec{j}  = - \frac{\nabla \psi \cdot (D \nabla C) }{\left| \nabla \psi \right|} = - D \frac{\partial C}{\partial n}= -B_{N}, 
\end{equation}
where $\vec{n} = \nabla \psi/|\nabla \psi|$ is the unit inward normal vector at the boundaries defined by the diffuse interface description. Equation \eqref{NBC1} can be rearranged to be $\nabla \psi \cdot (D \nabla C) = \left| \nabla \psi \right| B_{N}$ and  substituted back into Eq.~\eqref{MGE6}; thus, we obtain
\begin{equation} \label{MGE7}
\psi \frac{\partial C}{\partial t} = \nabla \cdot (\psi D \nabla C) - \left| \nabla \psi \right| B_{N} + \psi S. 
\end{equation}

To demonstrate that this smoothed boundary diffusion equation satisfies the assigned Neumann boundary condition (or specifying the boundary flux or normal gradient), we use the one-dimensional version of Eq.~\eqref{MGE7} without loss of generality. By reorganizing terms and integrating over the interfacial region, we obtain
\begin{equation} \label{SBM-Nm-prf-01}
\int_{a_i-\xi/2}^{a_i+\xi/2}\psi \left( \frac{\partial C}{\partial t} - S \right) dx = \left. \psi D \frac{\partial C}{\partial x} \right|_{a_i - \xi/2}^{a_i + \xi/2} - \int_{a_i - \xi/2}^{a_i + \xi/2} \left| \frac{\partial \psi}{\partial x} \right| B_{N} dx,
\end{equation} 
where $a_i-\xi/2 < x < a_i+\xi/2$ is the region of the interface, and $\xi$ is the thickness of the interface. Following Refs.~\cite{Bueno-Orovio:2006b,Bueno-Orovio:2006c,Kockelkoren:2003,Buzzard:2007}, we shall introduce the mean value theorem of integrals, which states that, for a continuous function, $g(x)$, there exists a constant value, $h_0$, such that:
\begin{equation}
\min{g(x)} < \frac{1}{q-p} \int_{p}^{q} g(x) dx = h_0 < \max{g(x)},
\end{equation}
where $p<x<q$. By eliminating the second term on the right-hand side of Eqs.~\eqref{MGE7}  and \eqref{SBM-Nm-prf-01},  the no-flux boundary condition can be imposed; the resulting equation is similar to those proposed in Refs. \cite{Bueno-Orovio:2006b,Bueno-Orovio:2006c,Kockelkoren:2003,Buzzard:2007}. However, we retain the term in order to maintain the generality of the method.  Therefore, the analysis presented here leads to an extension of the original method that greatly expands its applicability.

Since the function on the left-hand side of Eq.~\eqref{SBM-Nm-prf-01} is continuous and finite within the interfacial region, we can use the mean value theorem of integrals to obtain the relation:
\begin{equation}
\int_{a_i-\xi/2}^{a_i+\xi/2}\psi \left( \frac{\partial C}{\partial t} - S \right) dx = h_0 \xi. \label{SmVal-01}
\end{equation}
Using the conditions that $\psi = 1$ at $x=a_i + \xi/2$ and $\psi = 0$ at $x=a_i - \xi/2$, the first term in the right-hand side of Eq.~\eqref{SBM-Nm-prf-01} is written as
\begin{equation}\label{FLX-NM-01}
\left. 1 \cdot D \frac{\partial C}{\partial x} \right|_{a_i + \xi/2} - \left. 0 \cdot D \frac{\partial C}{\partial x} \right|_{a_i - \xi/2} = \left. D \frac{\partial C}{\partial x} \right|_{a_i + \xi/2}. 
\end{equation}
Since $| \partial \psi / \partial x | = 0$ for $x < a_i - \xi/2$ or $x > a_i+\xi/2$, the second term on the right-hand side of Eq.~\eqref{SBM-Nm-prf-01} can be replaced by
\begin{equation}\label{BC-NM-01}
\int_{a_i - \xi/2}^{a_i + \xi/2} \left| \frac{\partial \psi}{\partial x} \right| B_{N} dx = \int_{-\infty}^{+ \infty} \left| \frac{\partial \psi}{\partial x} \right| B_{N} dx. 
\end{equation}
Substituting Eqs.~\eqref{SmVal-01}, \eqref{FLX-NM-01} and \eqref{BC-NM-01} back into Eq.~\eqref{SBM-Nm-prf-01}, we obtain
\begin{equation} \label{SBM-Nm-prf-02}
h_0 \xi = \left. D \frac{\partial C}{\partial x} \right|_{a_i+\xi/2} - \int_{-\infty}^{+ \infty} \left| \frac{\partial \psi}{\partial x} \right| B_{N} dx. 
\end{equation}
Taking the limit of Eq.~\eqref{SBM-Nm-prf-02} for $\xi \rightarrow 0$, we obtain
\begin{equation} \label{SBM-Nm-prf-03}
\begin{split}
\left. D \frac{\partial C}{\partial x} \right|_{a_i} =  \int_{-\infty}^{+ \infty} \delta(x-a_i) B_N dx =  B_N \bigg|_{a_i},
\end{split}
\end{equation}
where $\partial C/\partial x|_{a_i+\xi/2} \cong \partial C/\partial x|_{a_i}$ and $\lim_{\xi \rightarrow 0} |\partial \psi / \partial x | = \delta(x-a_i)$ when $\psi$ takes the form of a hyperbolic tangent function, and $\delta(x-a_i)$ is the Dirac delta function. The Dirac delta function has the property that $\int_{-\infty}^{+\infty} \delta(x-a_i) f(x) dx = f(a_i)$, providing the second equality in Eq.~\eqref{SBM-Nm-prf-03}. Therefore, Eq.~\eqref{SBM-Nm-prf-03} clearly shows that the smoothed boundary method recovers the Neumann boundary condition at the boundary when the thickness of the diffuse boundary approaches zero. This convergence is satisfied for both stationary and moving boundaries \cite{Kockelkoren:2003}. 

For imposing the Dirichlet boundary condition, we can manipulate the original governing equation in a similar procedure to the derivation of Eq.~\eqref{MGE7}. Multiplying both sides of Eq.\ (\ref{MGE6}) with $\psi$, we obtain
\begin{equation} \label{MGE10}
\psi^{2} \frac{\partial C}{\partial t} = \psi \nabla \cdot ( \psi D \nabla C) - \psi \nabla \psi \cdot (D \nabla C) + \psi^{2} S, 
\end{equation}
where the second term on the right-hand side can be replaced by $\psi \nabla \psi \cdot( D \nabla C) = D [\nabla \psi \cdot \nabla \left( \psi C \right) - C \nabla \psi \cdot \nabla \psi] = D [\nabla \psi \cdot \nabla \left( \psi C \right) - C |\nabla \psi|^2]$. Equation \eqref{MGE10} is then rewritten as 
\begin{equation} \label{MGE11}
\psi^{2} \frac{\partial C}{\partial t} = \psi \nabla \cdot (\psi D \nabla C) - D[ \nabla \psi \cdot \nabla \left( \psi C \right) - C |\nabla \psi|^2] + \psi^{2} S,
\end{equation}
where $C$ in the third term will be the Dirichlet boundary condition, $B_D$, imposed at the diffuse interface. Therefore, the smoothed boundary formulated diffusion equation with the Dirichlet boundary condition is
\begin{equation} \label{MGE12}
\psi^{2} \frac{\partial C}{\partial t}  = \psi \nabla \cdot (\psi D \nabla C) - D[ \nabla \psi \cdot \nabla \left( \psi C \right) - B_D \left| \nabla \psi \right|^{2}] + \psi^{2} S.
\end{equation}

To prove the convergence of the solution at the boundaries to the specified boundary value, we again use a one-dimensional version of the smoothed boundary formulated equation. Integrating Eq.~\eqref{MGE12} over the interfacial region and reorganizing terms, we obtain
\begin{equation}\label{Int-Dirich-01}
\int_{a_i-\xi/2}^{a_i+\xi/2} \left[ \psi^2 \frac{\partial C}{\partial t} - \psi \frac{\partial}{\partial x} \left( \psi D \frac{\partial C}{\partial x} \right) - \psi^2 S \right] dx = -\int_{a_i-\xi/2}^{a_i+\xi/2} D \bigg(\frac{\partial \psi}{\partial x}\bigg) \left[ \frac{\partial \psi C}{\partial x}  - B_D \frac{\partial \psi}{\partial x} \right] dx.
\end{equation}
Similar to the derivation of Eq.~\eqref{SmVal-01}, the left-hand side of Eq.~\eqref{Int-Dirich-01} is proportional to the interfacial thickness and approaches zero in the limit of $\xi \rightarrow 0$. On the right-hand side of Eq.~\eqref{Int-Dirich-01}, the gradient of $\psi$ approaches the Dirac delta function, $\delta(x-a_i)$, as the interface thickness approaches zero. Therefore, we can reduce Eq.~\eqref{Int-Dirich-01} to 
\begin{equation}  \label{Int-Dirich-02}
0 =D \bigg[ \frac{\partial \psi C}{\partial x}  -  B_D \frac{\partial \psi}{\partial x}\bigg]  ~~~\Longrightarrow~~~ \frac{\partial \psi C}{\partial x} = B_D \frac{\partial \psi}{\partial x}
\end{equation}
in the limit $\xi \rightarrow 0$. By integrating over the interfacial region of Eq.~\eqref{Int-Dirich-02} again, we obtain
\begin{equation}
1 \cdot C \bigg|_{a_i+\xi/2} - 0 \cdot C \bigg|_{a_i-\xi/2} = \int_{a_i-\xi/2}^{a_i+\xi/2} B_D  \frac{\partial \psi}{\partial x} dx,\label{Int-Dirich-03}
\end{equation}
which in the limit of $\xi \rightarrow 0$ gives
\begin{equation}
C \bigg|_{a_i+\xi/2} \cong C \bigg|_{a_i} =\int_{-\infty}^{+ \infty} \delta(x-a_i) B_D dx = B_D \bigg|_{a_i}.
\end{equation}
Therefore, the smoothed boundary formulation recovers the specified Dirichlet boundary condition: $C|_{a_i} = B_D|_{a_i}$. 

In this method, the boundary gradient, $B_N$, and the boundary value, $B_D$, are not specified to be constant values. They can vary spatially and/or temporally or be functions of $C$ or other parameters. In addition, one can impose Neumann and Dirichlet boundary conditions simultaneously to yield mixed (or Robin) boundary conditions. The equation then becomes 
\begin{equation}\label{Ch6-SBM-02}
\psi^2 \frac{\partial C}{\partial t} = \psi \nabla \cdot (\psi D \nabla C) - \psi | \nabla \psi |_N B_N(\mathbf{x}) - \nabla \psi \cdot D[ \nabla (\psi C) - B_D(\mathbf{x})\nabla \psi ]_D+\psi^2 S,
\end{equation}
where $B_N(\mathbf{x})$ and $B_D(\mathbf{x})$ are spatially dependent Neumann and Dirichlet boundary conditions specified at different parts of the boundary, and the subscripts `$N$' and `$D$' denote the quantities associated with the boundaries to which the Neumann and Dirichlet boundary conditions are imposed.
 
\subsection{Surface Diffusion Coupled Bulk Diffusion} \label{SurfDiffFormulation}

The second example will demonstrate that surface diffusion can be implemented into the smoothed boundary equation derived above.  For this case, we take the set of equations that includes surface reaction, bulk diffusion and surface diffusion to describe an oxygen reduction model in a solid oxide fuel cell (SOFC) cathode \cite{Lu:2009}. The oxygen vacancy concentration, $C$, on the cathode surface is governed by Fick's Second Law:
\begin{equation} \label{FSL-S1}
D_b\frac{\partial C}{\partial n} = \kappa C -l D_s \bigg( \frac{\partial^2}{\partial s^2}+\frac{\partial^2}{\partial \tau^2} \bigg)C + L \frac{\partial C}{\partial t},
\end{equation}
where $n$, $s$ and $\tau$ are the unit normal, primary tangent and secondary tangent vectors of the surface, respectively. Here, the parameter $l$ is the characteristic thickness of the surface and is multiplied into the surface Laplacian term to maintain the dimensional agreement of the equation. The parameters $D_b$, $\kappa$, $D_s$, $L$ and $t$ are the bulk diffusivity, reaction rate, surface diffusivity, accumulation coefficient and time, respectively. Thus, the term on the left-hand side represents the flux from the bulk, and the terms on the right-hand side represent the surface reaction, surface Laplacian and concentration accumulation, respectively. For simplicity, these parameters are all assumed to be constant. In the bulk of cathode particles, the oxygen vacancy diffusion is also governed by Fick's Second Law:
\begin{equation} \label{FSL-B1}
\frac{\partial C}{\partial t} = D_b \nabla^2 C. 
\end{equation}
To simulate the oxygen vacancy concentration evolution in the cathode, the two diffusion equations, Eqs.~\eqref{FSL-S1} and \eqref{FSL-B1}, are coupled and need to be solved simultaneously. In this case, the two equations will share the flux normal to the cathode surface as the common boundary condition. Recently, this set of equations was formulated using the concept of diffuse interface approach \cite{Teigen:2009a}, which leads to two differential equations that are coupled by boundary conditions.  We will show below that the coupling can be achieved by applying the smooth boundary formulation described herein to obtain one single equation that governs both surface and bulk effects.

The derivation is as follows. We first multiply Eq.~\eqref{FSL-B1} with $\psi$ and applying the product rule of differentiation to obtain
\begin{equation} \label{SBM-FSL-B1}
\psi \frac{\partial C}{\partial t} = D_b\nabla \cdot (\psi \nabla C) - D_b \nabla \psi \cdot \nabla C.
\end{equation}
As in Eq.~\eqref{NBC1}, the normal derivative to the diffuse interface is defined by $\partial C/\partial n = \nabla C \cdot \nabla \psi /|\nabla \psi |$. Substituting this relation back into Eq.~\eqref{FSL-S1} and rearranging terms give
\begin{equation} \label{SBM-FSL-BC-1}
\nabla \psi \cdot \nabla C = \frac{|\nabla \psi|}{D_b} \bigg[ \kappa C - l D_s  \bigg( \frac{\partial^2}{\partial s^2}+\frac{\partial^2}{\partial \tau^2} \bigg)C + L \frac{\partial C}{\partial t} \bigg].
\end{equation}
Substituting Eq.~\eqref{SBM-FSL-BC-1} into the second term in Eq.~\eqref{SBM-FSL-B1}, we obtain
\begin{equation} \label{SBM-FSL-1}
\psi \frac{\partial C}{\partial t} =   D_b\nabla \cdot (\psi \nabla C) - |\nabla \psi| \bigg[ \kappa C- l D_s \bigg( \frac{\partial^2}{\partial s^2} +\frac{\partial^2}{\partial \tau^2} \bigg) C + L\frac{\partial C}{\partial t} \bigg].
\end{equation}
This equation combines the bulk diffusion and surface diffusion into one single equation, and will be used in examples presented in Sections \ref{bulkSurf_cylinder} and \ref{SOFC_Diff}. In the bulk ($|\nabla \psi|=0$ and $\psi = 1$), Eq.~\eqref{SBM-FSL-1} reduces back to Eq.~\eqref{FSL-B1}. When the interfacial thickness approaches zero, Eq.~\eqref{SBM-FSL-1} will reduce to Eq.~\eqref{FSL-S1} at the interface ($|\nabla\psi|\neq0$) as has been proven in Section \ref{DiffEqn}.

To calculate the surface Laplacian, we use the following method. The unit vector of the concentration gradient is given by $\vec{p} = \nabla C/|\nabla C|$. The unit secondary tangential vector on the surface can be obtained by $\vec{\tau} = (\vec{n} \times \vec{p})/|\vec{n} \times \vec{p}|$, and the unit primary tangential vector is then obtained by $\vec{s} = (\vec{\tau}\times \vec{n})/|\vec{\tau}\times \vec{n}|$. In this case, the surface flux has no projection in the $\tau$ direction ($\vec{p} \cdot \vec{\tau} = 0$). We can calculate the surface diffusion flux along the primary tangent direction simply by projecting the concentration gradient into the primary tangential direction. The surface flux is calculated by taking the inner product between the concentration gradient and the unit primary tangential vector for magnitude, and it is along the opposite primary tangential direction:
\begin{equation}
\vec{j}_s = - l D_s(\vec{p} \cdot \vec{s})\vec{s}.
\end{equation}
Since the Laplacian operator is independent of the selection of coordinate system, the value of the surface Laplacian can be then obtained by taking the negative divergence of the surface flux:
\begin{equation}
l D_s \bigg( \frac{\partial^2}{\partial s^2} +\frac{\partial^2}{\partial \tau^2} \bigg) C =  - \nabla \cdot \vec{j}_s,
\end{equation}
where $\nabla \cdot \vec{j}_s$ is the divergence of $\vec{j}_s$ in the global Cartesian grid system of the computational box.

To simulate only the surface diffusion on a diffuse-interface described geometry, one can simply eliminate all bulk-related terms to obtain
\begin{equation}
L\frac{\partial C}{\partial t} = - \nabla \cdot \vec{j}_s,
\end{equation}
such that only a concentration evolution along the interfacial region will occur.

\subsection{Mechanical Equilibrium  Equation}

The smoothed boundary method can also be applied to the mechanical equilibrium equation. When a solid body is in mechanical equilibrium, all the forces are balanced in all directions, as represented by
\begin{equation} \label{ME-1}
\frac{\partial \sigma_{ij}}{\partial x_j} = 0,
\end{equation}
where the subscript `$i$' indicates the component along the $i$-$th$ direction, and $\sigma_{ij}$ is the stress tensor. Repeated indices indicate summation over the index. For a linear elasticity problem, the stress tensor is given by the generalized Hooke's Law:
\begin{equation} \label{ESS-1}
\sigma_{ij} = C_{ijkl} (\varepsilon_{kl}- \rho \delta_{kl}),
\end{equation}
where $C_{ijkl}$ is the elastic constant tensor, and $\rho$ is a scalar body force, such as thermal expansion ($\alpha \Delta T$) or a misfit eigen-strain ($\varepsilon^0 = (a_p-a_m)/a_m$), which depends on the governing physics. The total strain tensor is defined by  the gradients of displacements as
\begin{equation} \label{TSN-1}
\varepsilon_{ij} = \frac{1}{2}\left(\frac{\partial u_i}{\partial x_j}+\frac{\partial u_j}{\partial x_i} \right),
\end{equation}
where $u_i$ is the infinitesimal displacement along the $i$-$th$ direction. Substituting Eqs.~\eqref{TSN-1} and \eqref{ESS-1} back into Eq.~\eqref{ME-1} gives
\begin{equation} \label{ME-2}
\frac{\partial }{\partial x_j} C_{ijkl}  \frac{1}{2} \left(\frac{\partial u_k}{\partial x_l}+\frac{\partial u_l}{\partial x_k} \right) = \frac{\partial }{\partial x_j} \bigg( \rho C_{ijkl}\delta_{kl} \bigg).
\end{equation}
We can multiply Eq.~\eqref{ME-2} by the domain parameter that distinguishes the elastic solid region ($\psi = 1$) from the environment ($\psi=0$) to perform the smoothed boundary formulation. After collecting the terms associated with $\partial \psi/\partial x_j$ on one side of the equation, we obtain
\begin{equation} \label{SBM-ME-1}
\begin{split}
\frac{\partial}{\partial x_j} \left[ \psi C_{ijkl}  \frac{1}{2} \left(\frac{\partial u_k}{\partial x_l}+\frac{\partial u_l}{\partial x_k} \right) \right] - \left(\frac{\partial \psi}{\partial x_j}\right) \bigg\{ C_{ijkl}  
\bigg[ \frac{1}{2}\bigg(\frac{\partial u_k}{\partial x_l} +\frac{\partial u_l}{\partial x_k} \bigg)-\rho\delta_{kl} \bigg] \bigg\} \\ = \frac{\partial}{\partial x_j} \bigg( \psi \rho C_{ijkl}\delta_{kl} \bigg).
\end{split}
\end{equation}

The traction exerted on the solid surface is defined by $N_{i} = -\sigma_{ij}n_{j}$, where $n_j$ is the inward unit normal of the solid surface. We again use the definition of the inward unit normal of the boundary: $n_{i} = \nabla \psi /|\nabla \psi |$. (In the indicial notation, $ \partial \psi/\partial x_i = \nabla \psi$ and $\sqrt{(\partial \psi/\partial x_i)(\partial \psi/\partial x_i)} = |\nabla \psi|$.) Therefore, the traction force is given by
\begin{equation} \label{Trac-1}
N_{i} = -\bigg\{ C_{ijkl} \left[ \frac{1}{2}\left(\frac{\partial u_k}{\partial x_l}+\frac{\partial u_l}{\partial x_k} \right)-\rho\delta_{kl} \right] \bigg\} \left(\frac{\nabla \psi}{|\nabla \psi|} \right).
\end{equation}
Substituting Eq.~\eqref{Trac-1} back into Eq.~\eqref{SBM-ME-1} returns the smoothed boundary formulation of the mechanical equilibrium equation with a traction boundary condition on the solid surface:
\begin{equation} \label{SBM-ME-2}
\frac{\partial}{\partial x_j} \left[ \psi C_{ijkl} \frac{1}{2} \left(\frac{\partial u_k}{\partial x_l}+\frac{\partial u_l}{\partial x_k} \right) \right] + |\nabla \psi| N_{i} = \frac{\partial}{\partial x_j} \bigg( \psi \rho C_{ijkl}\delta_{kl} \bigg),
\end{equation}
where $\partial(\psi \rho C_{ijkl}\delta_{kl})/\partial x_j = \tilde{\rho}_i$ can be treated as an effective body force along the $i$-$th$ direction.

For linear elasticity problems with presribed displacements at the solid surface, one can perform the smoothed boundary formulation as in the derivation of the Dirichlet boundary condition in Section \ref{DiffEqn} by multiplying Eq.~\eqref{ME-2} by $\psi^2$ and using the product rule to obtain
\begin{equation}
\begin{split}
\psi\frac{\partial}{\partial x_j} \left[ \psi C_{ijkl} \frac{1}{2} \left(\frac{\partial u_k}{\partial x_l}+\frac{\partial u_l}{\partial x_k} \right) \right] & - \bigg\{ \bigg(\frac{\partial \psi}{\partial x_j}\bigg) \bigg[ C_{ijkl} \frac{1}{2}\bigg( \frac{\partial \psi u_k}{\partial x_l}+\frac{\partial \psi u_l}{\partial x_k} \bigg) \bigg] \\ - \bigg(\frac{\partial \psi}{\partial x_j} \bigg) C_{ijkl} \frac{1}{2} \bigg( u_k \frac{\partial \psi}{\partial x_l} & + u_l \frac{\partial \psi}{\partial x_k}\bigg) \bigg\} = \psi^2 \frac{\partial}{\partial x_j}\bigg(\rho C_{ijkl}\delta_{kl}\bigg),
\end{split}
\end{equation}
where the displacements $u_k$ and $u_l$ appearing in the third term on the left-hand side will be the boundary values of the displacements at the solid surface. An equivalent formulation for the mechanical equilibrium equation can also be obtained by asymptotic approach \cite{Voigt:2009}.

\subsection{Equations for Phase Transformations with Additional Boundaries} \label{ContactAngleFormulation}

Phase transformations affected by a mobile or immobile surface or other boundary are of importance in many materials processes including heterogeneous nucleation that takes place at material interfaces \cite{Granasy:2007,Warren:2009}. Maintaining a proper contact angle at the three-phase boundary (where the interface between the two phases meets the surface) is necessary in capturing the dynamics accurately, since the contact angle represents the difference between surface energies (tensions) of different phase boundaries. While there are previous works that developed a method to impose the contact-angle boundary condition \cite{Granasy:2007,Warren:2009} on sharp domain walls, here we show that a similar model with diffuse domain walls can be obtained simply by applying the approach described above. Below, we assume that the boundary is immobile, but this assumption can be easily removed by describing the evolution of the domain parameter as dictated by the physics of the system.

In the Allen-Cahn and Cahn-Hilliard equations of the phase field model, the total free energy has the following form \cite{Cahn:1958,Cahn:1959}:
\begin{equation} \label{eqTeng}
F = \int_\Omega \bigg [ f(\phi)+\frac{\epsilon^{2}}{2} \left | \nabla \phi \right \vert^{2}\bigg]  d\Omega ,
\end{equation}
where $\phi$ is referred to as the phase field or order parameter commonly used to define different phases, and $\epsilon$ is the gradient energy coefficient in the phase field model. We take the variational derivative according to Euler's equation:
\begin{equation} \label{eqVD}
\delta F = \int_\Omega  \bigg( \frac{\partial f}{\partial \phi} - \epsilon^{2} \nabla^{2} \phi \bigg) d\Omega+ \int_{\partial \Omega}  \bigg( \epsilon^{2} \nabla \phi \cdot \vec{n}\bigg) d \vec{A},
\end{equation}
where $\vec{n}$ is the unit normal vector to the domain boundary $\partial \Omega$.  The bulk chemical free energy, $f$, is a double-well function of $\phi$. (This can also be derived from the order parameter $\phi$ changing with local ``velocity" $\dot{\phi}$.) For an extremum of the functional $F$, $\delta F = 0$ must be satisfied. This requirement provides two conditions:
\begin{subequations}
\begin{equation}\label{eqBC1}
 \frac{\partial f}{\partial \phi} - \epsilon^{2} \nabla^{2} \phi = 0 \quad \text{in}~~ \Omega,
\end{equation}
\begin{equation} \label{eqBC2}
 \epsilon^{2} \nabla \phi \cdot \vec n = 0 \quad \quad \text{on}~~ \partial \Omega.
\end{equation}
\end{subequations}
Following Eq.~\eqref{eqBC1}, we find  
\begin{equation} \label{eqSBC1}
\frac{\partial f}{\partial \phi} \nabla \phi = \epsilon^{2} \nabla^{2} \phi \nabla \phi = \frac{\epsilon^{2}}{2} \nabla (\left | \nabla \phi \right \vert^{2}),
\end{equation}
which can be rewritten as $\nabla f = \nabla ( \epsilon^2 | \nabla \phi |^{2})/2$. We thus find a useful equality for deriving the contact angle boundary condition:
\begin{equation}     \label{eqSBC2}
\left | \nabla \phi \right \vert = \frac{\sqrt{2 f}}{\epsilon}.
\end{equation}

In the smoothed boundary method, we introduce a domain parameter $\psi$ to incorporate boundary conditions in the original governing equation. As mentioned earlier, the level sets of this domain parameter $\psi$ describe the original boundaries and should satisfy $\vec{n} = \nabla \psi/| \nabla \psi |$. On $\partial \Omega$, we impose a contact angle $\theta$. Thus,
\begin{equation} \label{eqAngBC}
\vec n \cdot \frac{\nabla \phi}{\left | \nabla \phi \right \vert}  = \frac{\nabla \psi}{|\nabla \psi|} \cdot \frac{\nabla \phi}{|\nabla \phi|}= \cos \theta.
\end{equation}
Substituting Eq.~\eqref{eqSBC2} into Eq.~\eqref{eqAngBC}, one derives the following boundary condition formulation:
\begin{equation}  \label{eqAngBCF}
\nabla \psi \cdot \nabla \phi =  \left | \nabla \psi \right \vert \cos \theta \frac{\sqrt{2f}}{\epsilon}.
\end{equation}
This contact-angle boundary condition is similar to the one suggested by Warren et al.~\cite{Warren:2009} for contacting a sharp interface, for which a Dirac delta function will replace $|\nabla \psi |$.
 
The bulk chemical potential is defined by the variational derivative of the total free energy of the system:
\begin{equation} \label{mu-1}
\mu = \frac{\delta F}{\delta \phi} = \frac{\partial f}{\partial \phi} - \epsilon^2 \nabla^2 \phi,
\end{equation}
as it appeared in the first term of Eq.~\eqref{eqVD}. Multiplying both sides of Eq.~\eqref{mu-1} by the domain parameter $\psi$ gives
\begin{equation} \label{SBM-mu-1}
\psi \mu = \psi \frac{\partial f}{\partial \phi} - \psi \epsilon^2 \nabla^2 \phi = \psi \frac{\partial f}{\partial \phi} - \epsilon^2 \nabla \cdot ( \psi \nabla \phi ) + \epsilon^2 \nabla \psi \cdot \nabla \phi.
\end{equation}
We substitute the contact-angle boundary condition, Eq.~\eqref{eqAngBCF}, into the third term in Eq.~\eqref{SBM-mu-1} and obtain the smoothed boundary formulation for the chemical potential by dividing both sides by $\psi$:
\begin{equation}
\mu = \frac{\partial f}{\partial \phi} - \frac{\epsilon^2}{\psi} \nabla \cdot (\psi \nabla \phi) + \frac{\epsilon |\nabla \psi |}{\psi} \sqrt{2f} \cos{\theta}.
\end{equation}

For a nonconserved order parameter in the phase field models, the evolution is governed by the Allen-Cahn equation \cite{Allen:1979}, in which the order parameter evolves according to the local chemical potential variation:
\begin{equation} \label{AC-1}
\frac{\partial \phi}{\partial t} = - M \mu = -M \bigg( \frac{\partial f}{\partial \phi} - \frac{\epsilon^2}{\psi} \nabla \cdot (\psi \nabla \phi) + \frac{\epsilon |\nabla \psi |}{\psi} \sqrt{2f} \cos{\theta} \bigg).
\end{equation}
For a conserved order parameter, the evolution of the order parameter is governed by the divergence of the order-parameter flux, while the flux is proportional to the gradient of the chemical potential gradient. This process is governed by the Cahn-Hilliard equation \cite{Cahn:1958,Cahn:1959}:
\begin{equation} \label{CH-1}
\frac{\partial \phi}{\partial t} = \nabla \cdot (M \nabla \mu),
\end{equation}
for which the smoothed boundary formulation is obtained by (see Section \ref{DiffEqn}) 
\begin{equation} \label{SBM-CH-1}
\psi \frac{\partial \phi}{\partial t} = \nabla \cdot (\psi M \nabla \mu) - \nabla \psi \cdot (M \nabla \mu).
\end{equation}
Note that $-M \nabla \mu = \vec{j}$ is the flux of the conserved field order parameter. Therefore, the second term represents the fluxes normal to the domain wall (equivalent to Eq.~\eqref{NBC1}):
\begin{equation}
\nabla \psi \cdot (M \nabla \mu) = -(\vec{j}\cdot \vec{n}) |\nabla \psi |.
\end{equation}
Substituting the flux across the domain wall, our final smoothed boundary formulation of the Cahn-Hilliard equation is then written as 
\begin{equation}
\psi \frac{\partial \phi}{\partial t} = M \nabla \cdot \bigg[ \psi \nabla \bigg( \frac{\partial f}{\partial \phi} - \frac{\epsilon^2}{\psi} \nabla \cdot (\psi \nabla \phi) + \frac{\epsilon| \nabla \psi |}{\psi} \sqrt{2f} \cos{\theta} \bigg) \bigg] + |\nabla \psi |J_n,
\label{SBM-CH-3}
\end{equation}
where $J_n = \vec{j}\cdot \vec{n}$.  In practice, $\psi$ has a very small cutoff value such that the terms containing $1/\psi$ can be numerically evaluated. For time dependent problems, the equation is divided by $\psi$ before numerical implementation.

\section{Validation of the approach}

We demonstrate the validity and accuracy of the approach using bulk/surface diffusion in Sections \ref{DiffEqn} and \ref{SurfDiffFormulation}, as well as the phase transformation of three phase systems in Section \ref{ContactAngleFormulation}.

\subsection{1D Diffusion Equation}

First, we perform a 1D simulation to demonstrate that the Neumann and Dirichlet boundary conditions are satisfied on two different sides of the domain. Fick's second diffusion equation, Eq.~\eqref{OGE2}, with the given source term is solved within the solid phase that is defined by $\psi=1$. The diffusion coefficient $D$ is set to be 1, and the source $S$ is 0.02. On the right boundary of the solid, the gradient of $C$ is set to be -0.05, while on the left boundary, the value of $C$ is set to be 0.4. We perform the smoothed boundary formulation, as in the derivation of Eq.~\eqref{Ch6-SBM-02}, to obtain
\begin{equation} \label{SBM-Diff-A}
\psi^2 \frac{\partial C}{\partial t} = \psi \nabla \cdot (\psi\nabla C) - \psi [| \nabla \psi | (-0.05)]_r -[\nabla \psi \cdot \nabla (\psi C) - |\nabla \psi|^2(0.4)]_l +\psi^2 (0.02),
\end{equation}
where the subscripts `$r$' and `$l$' indicate the right and left interfaces. The solid region is approximately in the range between the 102-$th$ and 298-$th$ grid points. We use a hyperbolic tangent form for the continuous domain parameter $\psi$ as
\begin{equation}
\psi = \frac{1}{2}\{\tanh{[0.8(x-10)+1]}-\tanh{[0.8(x-30)+1]}\},
\end{equation}
such that the interfacial thickness is taken to be approximately 6 grid spacings. The initial concentration is $C=0$ everywhere in the computational box. A standard finite central differencing scheme in space and the Euler explicit time scheme are employed in the simulation. The grid spacing is taken to be $\Delta x = 0.1$.

\begin{figure}[htb] 
\begin{center}
\includegraphics[width=1\textwidth]{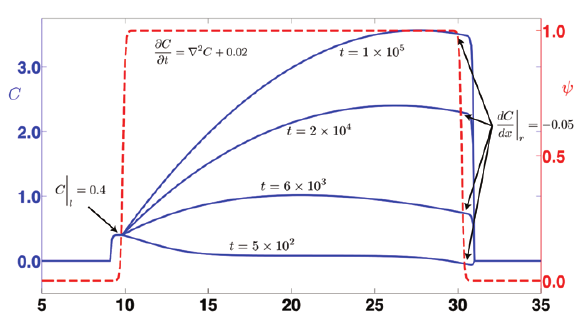}
\end{center}
\caption{Demonstration of the smoothed boundary method on the 1D diffusion equation. The red dashed line is the domain parameter, and the blue lines are the concentration profiles taken at different times. The Neumann BC is imposed at the right boundary, and the Dirichlet BC is imposed at the left boundary.}
\label{1D_demo}
\end{figure}

Figure \ref{1D_demo} shows the concentration profiles taken at four different times (in blue solid lines). The domain parameter is plotted in the red dashed line. On the right interface, it can be clearly observed that $dC/dx = -0.05$ at all times, except for a rapid change from $dC/dx=0$ to $dC/dx = -0.05$ in the very early transient period. In the early period, the concentration even takes negative values to satisfy the gradient boundary condition imposed at the right boundary. On the other hand, the concentration remains at 0.4 during the entire diffusion process, except in the very early transient period during which $C$ changed from 0 to 0.4. This result clearly demonstrates both Neumann and Dirichlet boundary conditions are satisfied on the diffuse interfaces.

\subsection{Surface Diffusion and Bulk Diffusion in a Cylinder} \label{bulkSurf_cylinder}

To further demonstrate the validity of the smoothed boundary method, we apply the method to a cylinder for which a cylindrical coordinate grid system can be used. We solve the coupled surface-bulk diffusion problem using both the smoothed boundary and standard sharp interface formulations in the same grid system for comparison. Again, we use a continuous domain parameter $\psi$ to define the solid region of a cylinder ($\psi = 1$ for solid, and $\psi=0$ for environment). The solid surface is then represented by $0<\psi<1$. 

For the smoothed boundary case, we solve Eq.~\eqref{SBM-FSL-1} using the central differencing scheme in space. The radial direction is discretized into 80 grid points, and the longitudinal direction is discretized into 600 grid points. The grid spacing $\Delta x$ is $1/60$, such that the radius of the cylinder is $R = 60\Delta x = 1$, and the length of the cylinder is $10R$. The thickness of the diffuse interface is approximately 4$\sim$5 grid spacings; thus, the characteristic thickness appearing in Eq.~\eqref{SBM-FSL-1} is set to be $l= 4.5\Delta x = 0.075$. Here, we set the surface accumulation coefficient $L$ to be 0 for simplicity. We investigate two cases: one with a low surface reaction rate, $\kappa = 2.1$, and the other with a high surface reaction rate, $\kappa = 1000$.

To compare the results, we solve the original form of the coupled surface and bulk diffusion equations using the sharp interface approach with the same finite difference method. The same grid system is used, except a cylinder surface is now explicitly placed at $R =1$ at which the boundary condition is imposed. In this case, we calculate the normal flux to the surface by the right-hand side of Eq.~\eqref{FSL-S1} with the grid system on the cylinder surface, and then use the flux as the boundary condition for the concentration evolution in the bulk and on the surface, Eq.~\eqref{FSL-B1}. Note that the characteristic thickness $l$ drops in the sharp interface description as the limit $l \rightarrow 0$ is taken.

\begin{figure}[htb]
\begin{center}
\includegraphics[width=1\textwidth]{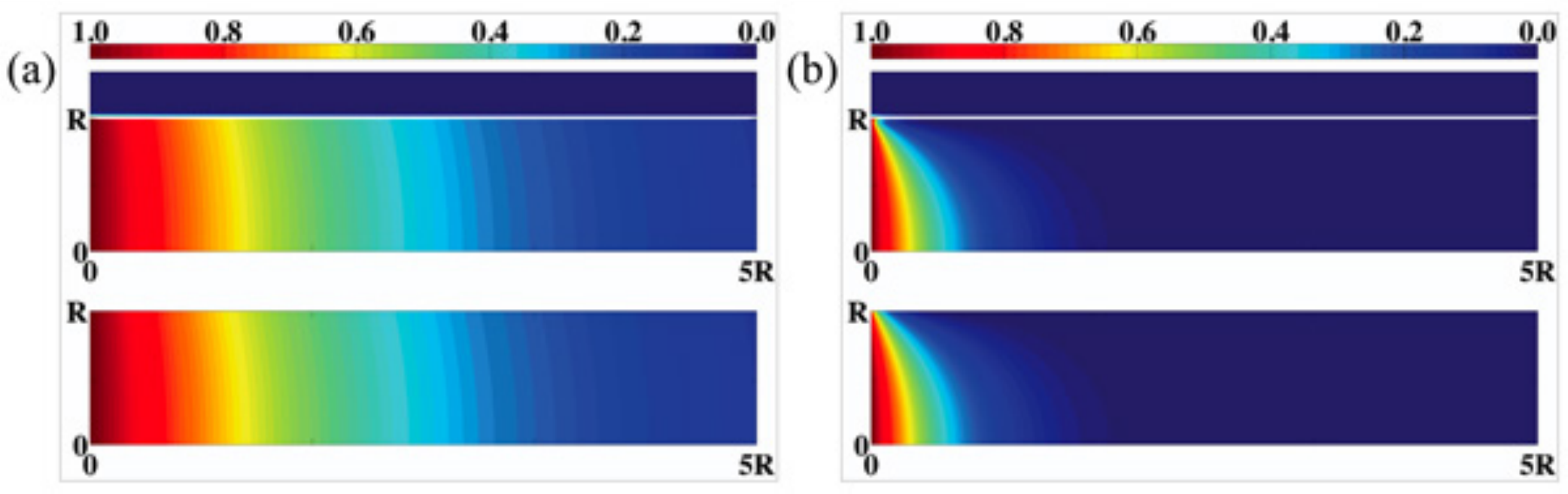}
\caption{The concentration profiles (a) for $D_b=1$, $\kappa=2.1$, $D_s = 10$ and (b) for $D_b=1$, $\kappa=1000$, $D_s = 10$ obtained by the smoothed boundary method (top) and the finite difference method with sharp interface model (bottom). The top region with constant blue color is outside of the solid, while the solid white lines indicate the solid surface.}\label{Cyl-Con}
\end{center}
\end{figure}

Figures \ref{Cyl-Con}(a) and (b) show the concentration profiles in the cylinder at the steady state obtained using the smoothed boundary method and the finite central difference method. For clarity, only the concentration in the region of $0<z<5R$ is presented. The top rows in Figs.~\ref{Cyl-Con}(a) and (b) are the smoothed boundary results, and the bottom rows are the sharp interface results. The results from the two methods are clearly in excellent agreement.

\begin{figure}[htb]
\begin{center}
\includegraphics[width=1\textwidth]{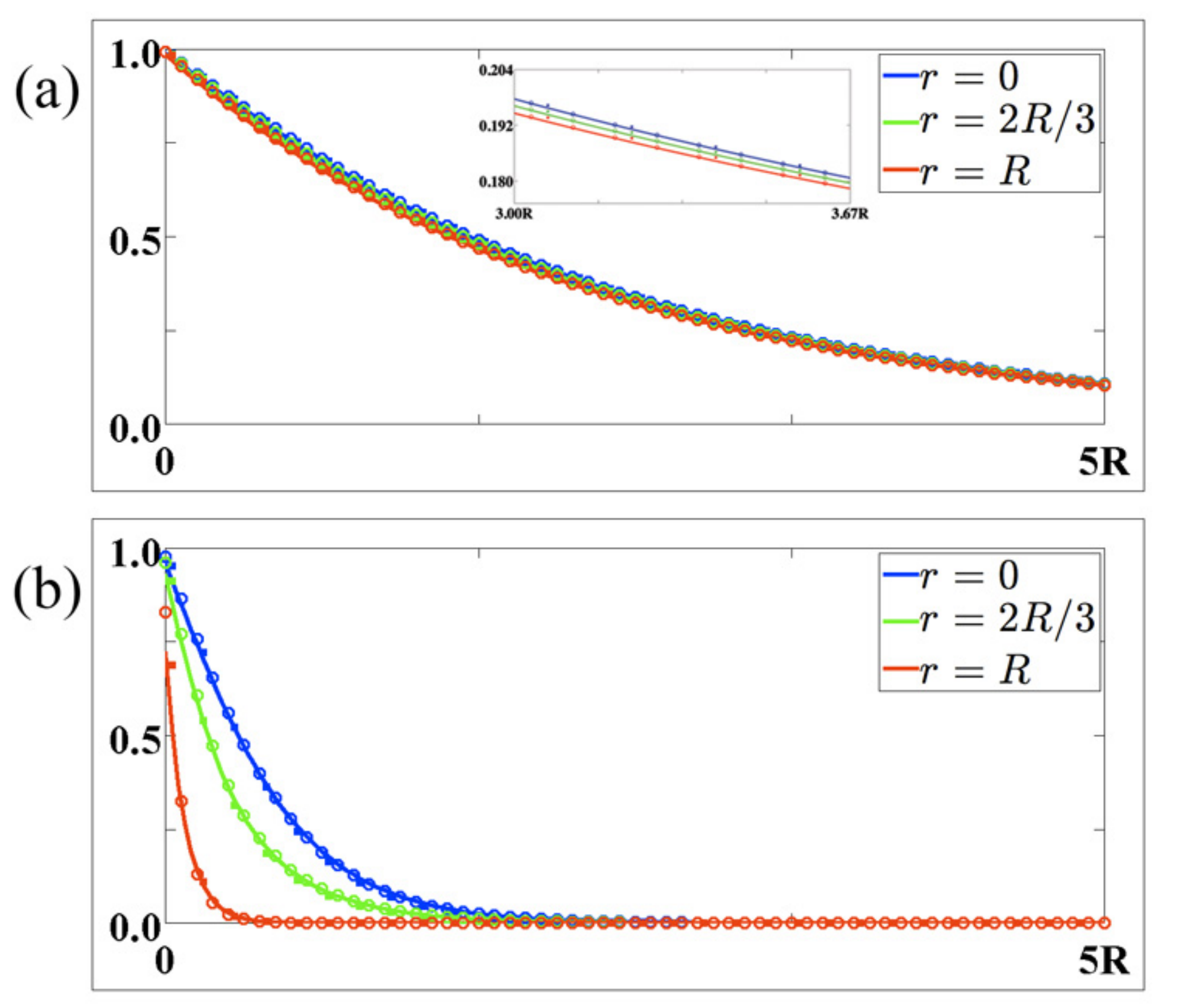}
\caption{The concentration along the line at $r=0$, $r=2R/3$ and $r=R$ for (a) $D_b=1$, $\kappa=2.1$ and $D_s = 10$; and (b) $D_b=1$, $\kappa=1000$ and $D_s = 10$. The solid lines are the solutions from the finite difference method with sharp interface model. The circular markers are the solutions from the smoothed boundary method with 4.5 grid spacings across the interface, $\Delta x = 1/60$ and $R=60\Delta x=1$. The solid square markers are the smoothed boundary solutions with 4.5 grid spacings across interface, $\Delta x = 1/30$ and $R=30\Delta x=1$. To clearly illustrate the concentration profile for the low surface reaction case, a magnified view is provided in (a). }\label{Cyl-Err}
\end{center}
\end{figure}

Shown in Figs.~\ref{Cyl-Err}(a) and (b) are the concentration profiles plotted along the longitudinal line at $r=0$, $r =2R/3$, and $r = R$, respectively. Again the plots show that the differences between the results from the two methods are small for the cylindrical geometry. As mentioned in a previous section, the error of the smoothed boundary method is proportional to the interfacial thickness. Based on our tests, we found that, even for an interfacial-thickness-to-radius ratio of around 1/5, the maximum error between the two methods appearing near the surface is still around 2\% (shown in the solid square markers), while the error in the bulk region is significantly smaller than that. If we select the interfacial-thickness-to-radius ratio to be 1/10, the maximum error appearing in the entire solid region is on the order of $1\times10^{-3}$ (including the region near the surface). Another controlling factor of errors is the number of grid points across the diffuse interface. From our numerical tests, we noticed that at least 4 grid points are required to properly resolve the sharp change in $\psi$ across the interface such that the errors are reasonably small. In addition to the steady state solution, the transient state solutions are also in excellent agreement during the entire diffusion process. This demonstrates that the smoothed boundary method can be employed to accurately solve coupled surface diffusion and bulk diffusion problems.

\subsection{Contact Angle Boundary Condition}

We perform a simple 2D simulation to validate the smoothed boundary formulation for the contact-angle boundary condition at the three-phase boundary. Equations \eqref{AC-1} and \eqref{CH-1} are tested for nonconserved and conserved field order parameters, respectively. The computational box sizes are $L_x= 100$ and $L_y=100$, and the parameters used are $\Delta x= 1$, $M= 1$, and $\epsilon = 1$. On the computational box boundaries, the normal gradients of the order parameter are set to be zero: $\partial \phi/\partial x = 0$ at $x = 0$ and $x=100$ and $\partial \phi/\partial y =0$ at $y=0$ and $y=100$. A horizontal flat wall is defined by a hyperbolic tangent function of the domain parameter $\psi$ 
\begin{equation}
\psi = \frac{1}{2}\tanh{(y-30)}+\frac{1}{2},
\end{equation}
such that $\psi=0.5$ is at $y=30$ and $\psi$ gradually transitions from 0 to 1 from below the wall to above. The wall thickness is approximately 5 grid spacings. The initial phase boundary is vertically placed at the middle of the domain ($x=50$) with phase 1 ($\psi=1$) and phase 0 ($\psi=0$) on the left and right halves, respectively.

\begin{figure}[htb]
\begin{center}
\includegraphics[width=1\textwidth]{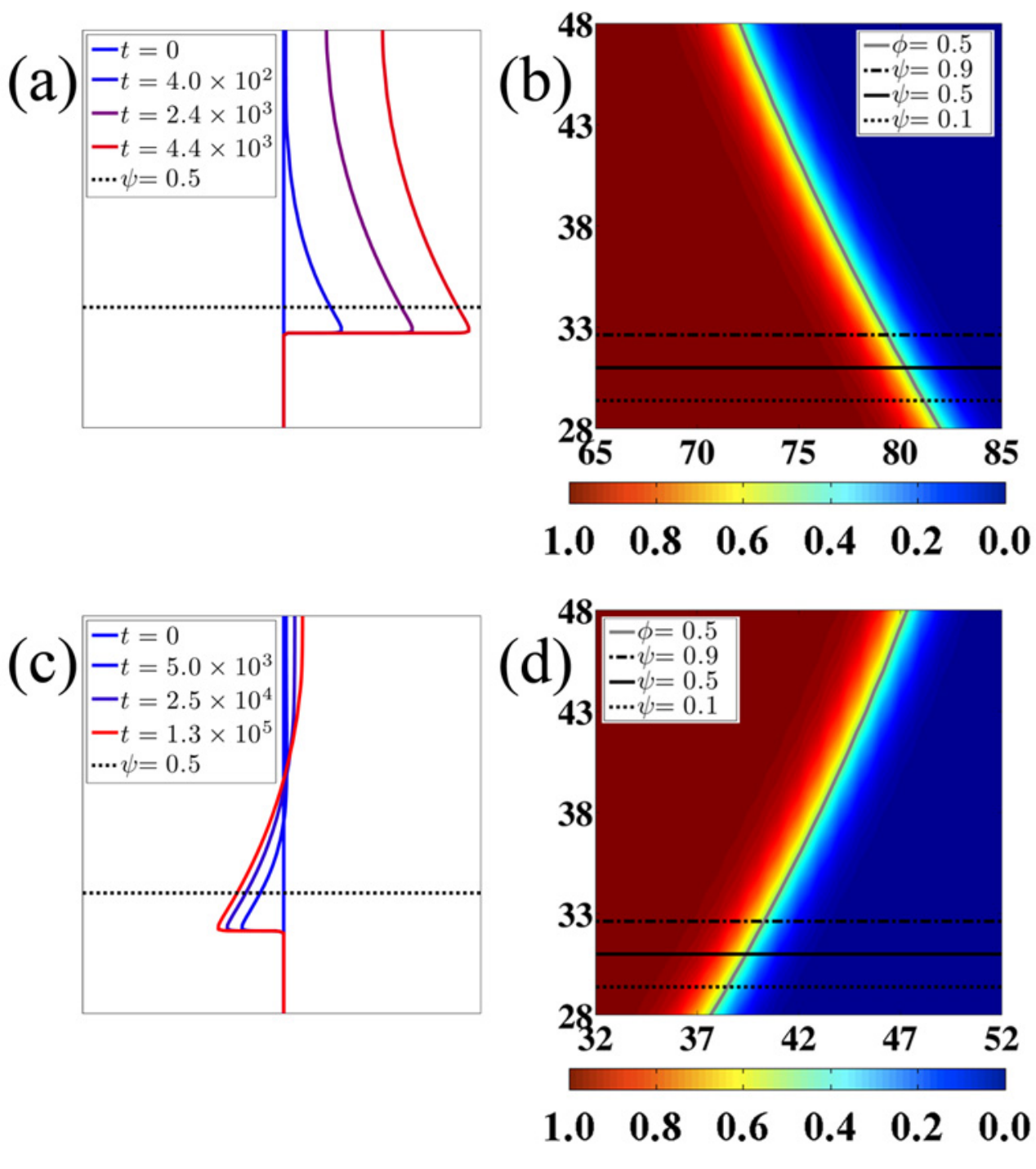}
\caption{(a) Allen-Cahn type phase transformation with a 60$^\circ$ contact-angle BC. (b) Magnified view of the order parameter profile at the three-phase boundary, corresponding to $t= 2.4\times10^{3}$ in (a). (c) Cahn-Hilliard type phase transformation with a 120$^\circ$ contact-angle BC. (d) Magnified view of the order parameter profile at the three-phase boundary, corresponding to $t= 1.3\times10^{5}$ in (c). The imposed contact angles can be clearly verified in (b) and (d). The field order parameters in the region of $\psi < 0.5$ have no physical significance. For Cahn-Hilliard case, the field order parameter is conserved in the region of $\psi>0.5$.}\label{CA-Validate}
\end{center}
\end{figure}

In the first case with nonconserved order parameter, we evolve Eq.~\eqref{AC-1} with a 60-degree contact angle. The result clearly shows a 60-degree contact angle at the three-phase boundary as imposed (Fig.~\ref{CA-Validate}(a)). The angle can be measured by the intersection between the two contours of $\psi = 0.5$ and $\phi=0.5$, as shown in Fig.~\ref{CA-Validate}(b). The 60-degree angle is maintained during the entire evolution, except for the very early transient period when the contact angle changed from 90 to 60 degrees. Due to the imposed contact angle, the initially flat phase boundary bends and creates a negative curvature of phase 1. As a result, the phase boundary moves toward phase 0, and eventually only phase 1 remains in the system.

For the second case with conserved order parameter, we evolve Eq.~\eqref{CH-1} with a contact angle of 120 degrees. As expected, the phase boundary intersects the wall at a 120-degree contact angle (Fig.~\ref{CA-Validate}(c) and (d)). In contrast to the Allen-Cahn type dynamics, due to the conservation of the order parameter, the phase boundary near the wall moves toward the left while the phase boundary away from the wall moves in the opposite direction. As a result, the phase boundary deforms to a curved shape. When the system reaches its equilibrium state, the phase boundary forms a circular arc with a uniform curvature everywhere along the phase boundary, such that the total surface energy is minimized (see Fig.~\ref{CA-Validate}(c) for $t=1.3\times10^5$). 

\section{Applications}

While the details of the scientific calculations performed by applying of these methods will be published elsewhere, it is worthwhile to show some of the results to demonstrate the potential of the method.  

\subsection{Surface-Reaction Diffusion Kinetics} \label{SOFC_Diff}

The first example is ionic transport through a complex microstructure.  Here, the ion diffusion is driven by a sinusoidal voltage perturbation. For the steady state solution, the time dependence of the form $\exp(\mathrm{i} \omega t)$, where $\omega$ is the angular frequency and $\mathrm{i}=\sqrt{-1}$, can be removed, as in the equation derived by Lu et al.~\cite{Lu:2009}. For a demonstration, we solve the steady state solution for the case without surface diffusion while solving the transient state solution for the case with surface diffusion. For the first case, the smoothed boundary formulated equation is given by
\begin{equation} \label{SBM-Cx-1}
\nabla \cdot (\psi \nabla \tilde{C}) - |\nabla \psi|\kappa \tilde{C} = \mathrm{i} \omega \psi \tilde{C},
\end{equation}
where $\tilde{C}$ is the concentration amplitude consisting of a real and imaginary part. This equation is solved by a standard alternative direction iterative (ADI) method in a second-order central-difference scheme in space ($\Delta x= 0.04$). For the transient state solution, we keep the time dependence as is, and the smoothed boundary formulation is given by Eq.~\eqref{SBM-FSL-1} in which surface diffusion, bulk diffusion and surface reactions are all considered. For simplicity, the surface accumulation   term is ignored ($L=0$). Equation \eqref{SBM-FSL-1} is solved by a second-order central-difference scheme in space ($\Delta x= 0.04$) and the Euler explicit time stepping scheme ($\Delta t = 0.01$). Here, we employed an Allen-Cahn type equation \cite{Beckermann:1999,Jamet:2008a,Jamet:2008b} to smooth the initially sharp boundaries of experimentally obtained 3D voxelated data ($\psi =1$ for the voxels in the cathode and $\psi=0$ for the voxels in the pores):
\begin{equation} \label{S-1}
\frac{\partial \psi}{\partial t} = -\frac{\partial f}{\partial \psi}+\epsilon^2 \nabla^2 \psi - \epsilon\sqrt{2f}\frac{|\nabla \psi | \nabla^2 \psi - \nabla \psi \cdot \nabla |\nabla \psi |}{(\nabla \psi)^2}\chi,
\end{equation}
where $f = \psi^2(1-\psi)^2$ is a typical double-well function, and $\epsilon$ is the gradient energy coefficient. The interfacial thickness ($0.1<\psi<0.9$) is given by $2\epsilon\sqrt{2}$. Note that the third term in Eq.~\eqref{S-1} is used to remove the curvature effect such that the location of $\psi = 0.5$ does not change during the smoothing process if $\chi=1$. The computational box contains 321, 160 and 149 grid points along the $x$, $y$ and $z$ directions. 

\begin{figure}[htb]
\begin{center}
\includegraphics[width=1\textwidth]{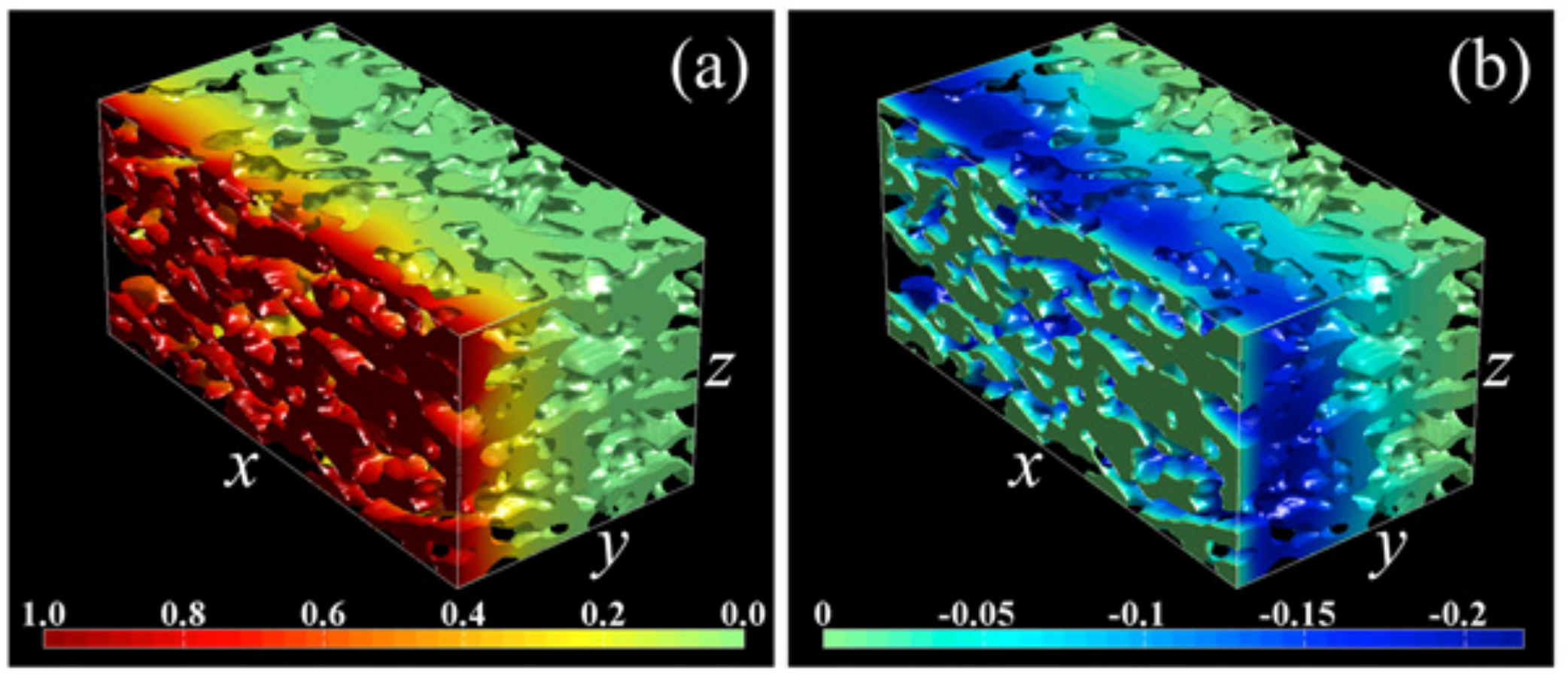}
\caption{Steady state concentration for $D_b = 1$, $\kappa = 0.1$ and $D_s= 0$ in a real cathode complex microstructure: real part (a) and imaginary part (b).}\label{CRI-AC-1}
\end{center}
\end{figure}

Figure \ref{CRI-AC-1} shows the steady-state concentration for the case in which $D_b = 1$, $\kappa = 0.1$, $D_s= 0$ and $\omega=0.55$. The boundary condition on the computational box is $\tilde{C} = 1$ at $y=0$, $\tilde{C} = 0$ at $y=6.4$ and no-gradient on the remaing four sides. As shown in Fig.~\ref{CRI-AC-1}(a), the real part of the concentration decays from 1 to 0 over the complex cathode microstructure to satisfy the boundary condition given on the domain box at $y=0$ and $y = 6.4$. For the imaginary part, the values at $y=0$ and $y=6.4$ remain at 0 as assigned. In the middle region, a negative value of the imaginary part occurs due to the phase shift resulting from the delayed response.

\begin{figure}[htb]
\begin{center}
\includegraphics[width=1\textwidth]{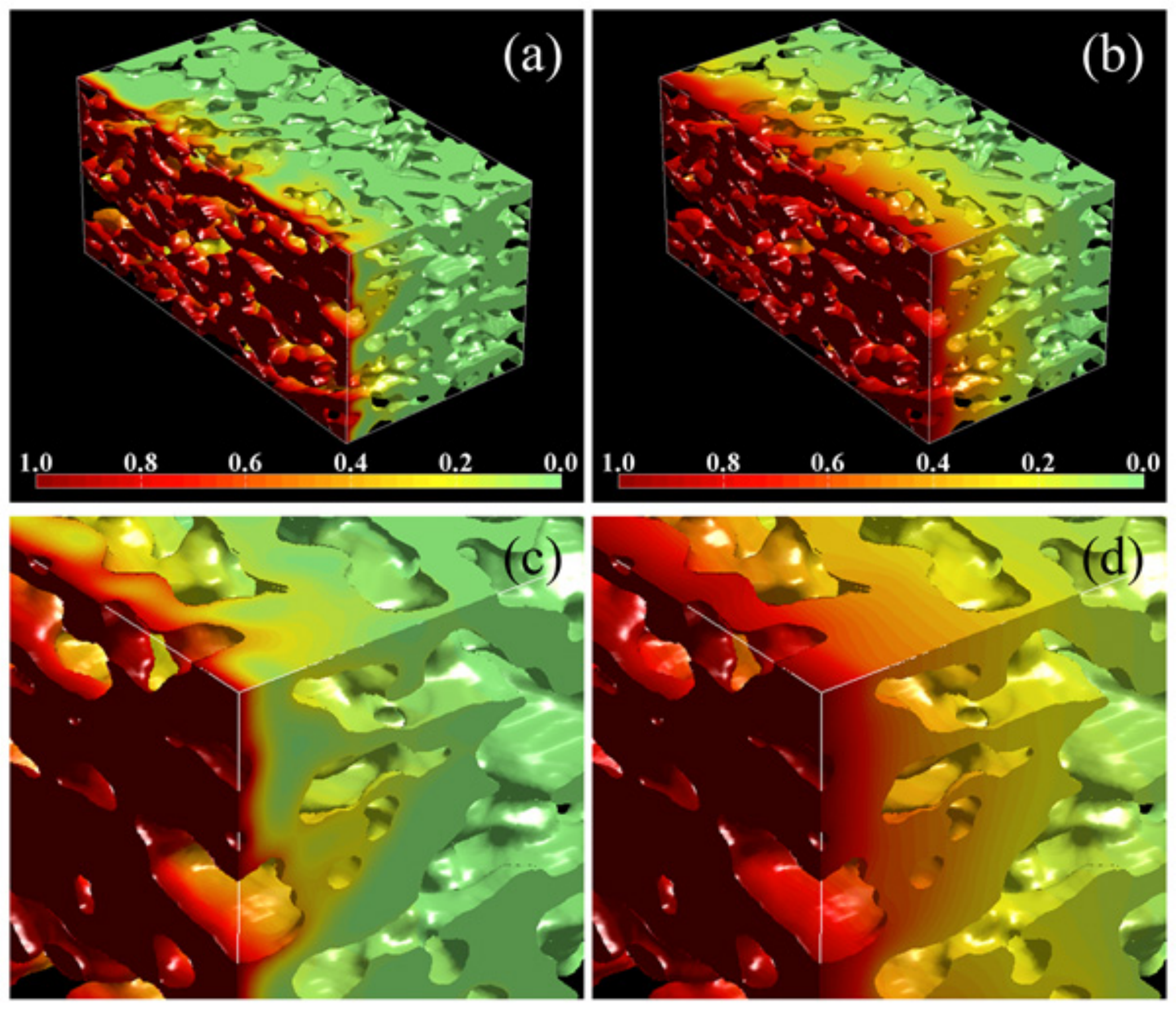}
\caption{The concentration for $D_b = 1$, $\kappa = 2.1$ and $D_s= 10$ at intermediate time (a) $t =2\times10^{-2}$, and (b) $t=4.9\times10^{-1}$ in a real cathode complex microstructure. Figures (c) and (d) are the magnified views of (a) and (b).}\label{SurfDiff_1}
\end{center}
\end{figure}

Figure \ref{SurfDiff_1} shows the concentration distribution taken at two different times for the case in which $D_b = 1$, $\kappa = 2.1$ and $D_s= 10$, with DC loading ($\omega=0$). The boundary conditions on the computational box are the same as in the AC loading case above. An enhanced concentration along the irregular surface due to surface diffusion can be clearly observed in the intermediate stage, Fig.~\ref{SurfDiff_1}(a). As the concentration propagates through the bulk region, the system eventually approaches its steady state, and the concentration enhancement diminishes, Fig.~\ref{SurfDiff_1}(b). Figures \ref{SurfDiff_1}(c) and (d) are magnified views of (a) and (b).

\begin{figure}[htb]
\begin{center}
\includegraphics[width=1\textwidth]{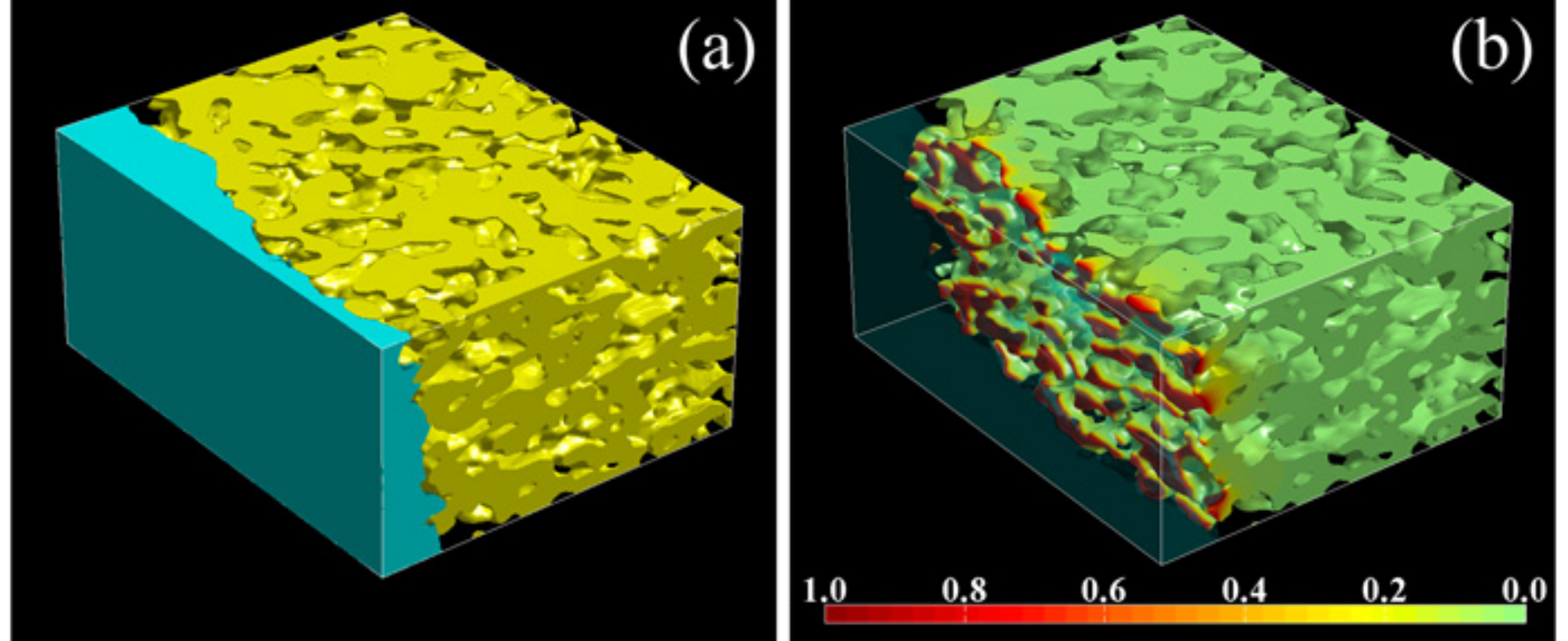}
\caption{(a) Microstructure of the electrolyte and cathode in SOFC, in which cyan and yellow colors indicate the electrolyte (YSZ) and cathode (LSC) phases, respectively. (b) The steady-state concentration in the complex cathode phase for $D_b = 1$, $\kappa = 0$ and $D_s= 0$ with fixed $C=1$ at the irregular LSC-YSZ contact surface. }\label{DiriBC}
\end{center}
\end{figure}

The smoothed boundary method can also be used to impose Dirichlet boundary conditions on irregular surfaces. For example, if the ion diffusivity is much higher in the electrolyte phase than in the cathode phase, the concentration in the electrolyte will be nearly uniform. To simulate this scenario, we impose a fixed concentration at the electrolyte-cathode contacting surface as the boundary condition. On the computational box boundaries, we set $C=0$ at $y=10.44$ and the no-flux boundary condition for the remaining five sides. The material parameters are selected to be $D_b= 1$, $\kappa=0$, and $D_s = 0$. Figure \ref{DiriBC} shows the simulation results for a pure bulk diffusion example with a fixed value $C=1$ imposed at the LSC (cathode) -YSZ (electrolyte) interfaces. In this case, since the contacting areas are small (compared to the cross-sectional area of LSC on the $x$-$z$ plane in Fig.~\ref{CRI-AC-1}(a)), ion diffusion along the lateral directions ($x$ and $z$) is large. As a result, the concentration drops very rapidly in a short distance from the contacting areas. Therefore, the concentration distribution is very different from the ones shown in Figs.~\ref{CRI-AC-1} and \ref{SurfDiff_1}, where the cross-sectional areas at $y=0$ and $y=6.4$ ($x$-$z$ planes on the computational box) are approximately equal. 

\subsection{Kirkendall Effect Diffusion with a Moving Boundary Driven by Coupled Navier-Stokes-Cahn-Hilliard equations} \label{Kirkendall effect deformation}

The third application will demonstrate the smoothed boundary method's broad applicability by applying it to the coupled Navier-Stokes-Cahn-Hilliard equations \cite{Gurtin:1996,Jacqmin:1999,Kim:2005,Zhou:2006,Villanueva:2008}. This particular formulation aims to solve diffusion problems with the Kirkendall effect with vacancy sources and sinks in the bulk of the solid \cite{Kirkendall:1939,Kirkendall:1942,Smigelskas:1947,Darken:1948,AtomMovements:Bardeen}. In this case, the solid experiences deformation due to vacancy generation and elimination. The Navier-Stokes-Cahn-Hilliard equations are coupled to the smoothed boundary formulation of the diffusion equation in Section \ref{DiffEqn} as a model of plastic deformation due to volume expansion and contraction resulting from vacancy flow.

When the diffusing species of a binary substitutional alloy have different mobilities, the diffusion fluxes of the two species are unbalanced, creating a net vacancy flux toward the fast diffuser side. Here, we denote the slow diffuser, fast diffuser and vacancy by $A$, $B$ and $V$, respectively. Due to the accommodation/supply of excess/depleted vacancies, the solid locally expands/shrinks \cite{Strandlund:2004,Larsson:2006,Strandlund:2006,Yu:2007} when maintaining the vacancy mole fraction at its thermal-equilibrium value. We treat the solid as a very viscous fluid \cite{Stephenson:1988,Boettinger:2005,Dantzig:2006,Boettinger:2007} with a much larger viscosity than the surrounding environment. In this case, we solve the Navier-Stokes-Cahn-Hilliard equations to update the shape of the material  as follows \cite{Yu:2009a}:
\begin{subequations}
\begin{equation} \label{NS-CH-1}
-\nabla P+\nabla \cdot (\eta \nabla \mathbf{v})+\nabla \bigg(\frac{2\eta}{d}S_V\bigg)+\frac{1}{C_a}\mu\nabla \psi=0,
\end{equation}
\begin{equation}  \label{NS-CH-2}
\nabla \cdot \mathbf{v}=-S_V,
\end{equation}
\begin{equation} \label{NS-CH-3}
\frac{\partial \psi}{\partial t} -\mathbf{v}\cdot \nabla \psi = M \nabla^2\bigg( \frac{\partial f}{\partial \psi} -\epsilon^2 \nabla^2\psi\bigg),
\end{equation}
\end{subequations}
where $P$ is the effective pressure, $\eta$ is the viscosity, $\mathbf{v}$ is the velocity vector, $d$ is the number of dimension, and $C_a$ is the Cahn number reflecting the capillary force compared to the pressure gradient. One great convenience of solving this type of phase field equation is that it automatically maintains the domain parameter in the form of a hyperbolic tangent function while updating the location of the diffuse interface. Note that we have ignored the inertial force in the Navier-Stokes equation to obtain Eq.~\eqref{NS-CH-1} since the deformation is assumed to be a quasi-steady state process. The vacancy generation rate that results in the local volume change is given by
\begin{equation}
S_V = -\frac{\nabla \cdot (D_{VB} \nabla X_B)}{\rho_l(1-X_V^{eq})},
\end{equation}
where $X_B$ is the mole fraction of the fast diffuser, $X_V^{eq}$ is the thermal-equilibrium vacancy mole fraction, $D_{VB}$ is the diffusivity for vacancy flux associated with $\nabla X_B$, and $\rho_l$ is the lattice site density. The fast diffuser mole fraction evolution is governed by the advective Fick's diffusion equation as
\begin{equation} \label{KE-Trad-1}
\frac{\partial X_B}{\partial t} -\mathbf{v} \cdot \nabla X_B= \nabla \cdot (D_{BB}^V\nabla X_B) -X_B S_V,
\end{equation}
where $D_{BB}^V$ is the diffusivity for a fast diffuser flux associated with $\nabla X_B$, and the advective term accounts for the lattice shift due to volume change. Since the diffusing species cannot depart the solid, a no-flux boundary condition is imposed at the solid surface. Thus, we obtain the smoothed boundary formulation of Eq.~\eqref{KE-Trad-1} as
\begin{equation} \label{KE-Trad-SBM-1}
\psi \bigg( \frac{\partial X_B}{\partial t} -\mathbf{v} \cdot \nabla X_B \bigg)= \nabla \cdot (\psi D_{BB}^V\nabla X_B) -\psi X_B S_V.
\end{equation}
As the concentration evolves, the shape of the solid is also updated by Eq.~\eqref{NS-CH-3} and iteratively solving Eqs.~\eqref{NS-CH-1} and \eqref{NS-CH-2} by applying a projection method \cite{Kim:2006a,Kim:2006b}.

\begin{figure}[htb]
\begin{center}
\includegraphics[width=1\textwidth]{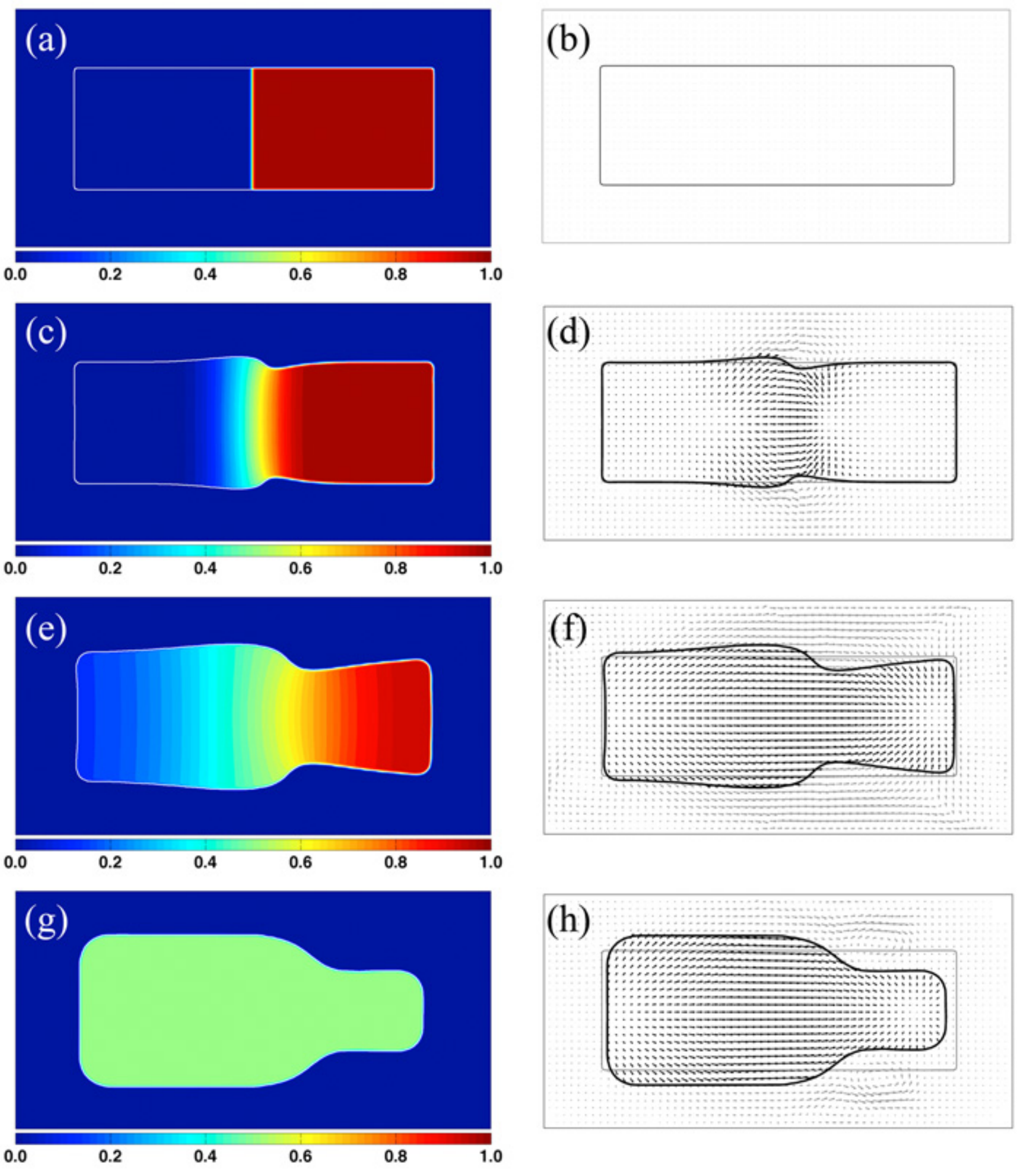}
\caption{ Left column: Normalized concentration profiles (to the lattice site density) taken at difference times. Right column: Velocity fields taken at different times corresponding to the left column. Black and gray arrows denote the flow inside and outside the material, respectively. The flow outside of the material has no physical significance to the shape change. }\label{Def_Con}
\end{center}
\end{figure}

The slow and fast diffusers are initially placed in the left and right halves of the solid, respectively. We use theoretically calculated diffusivities for this simulation \cite{Moleko:1989,Manning:1971,Belova:2000,VanDerVen:2009}. Figure \ref{Def_Con} shows snapshots of the concentration profiles (left column) and velocity fields (right column) from a 2D simulation. As the fast diffuser diffuses from the right to the left side, the vacancy elimination and generation cause contraction and expansion on the right and left sides, respectively. As a result, the initially rectangular slab deforms to a bottle-shaped object.

In another scenario in which the vacancy diffusion length is comparable to or smaller than the distance between vacancy sources and sinks, the explicit vacancy diffusion process must be considered \cite{VanDerVen:2009,Yu:2008}. In this case, vacancies diffuse in the same manner as the atomic species. In the bulk of a solid devoid of vacancy sources/sinks, the concentration evolutions are governed by
\begin{subequations} \label{KE-Rig-1}
\begin{equation}
\frac{\partial X_V}{\partial t} = \nabla \cdot (D_{VV} \nabla X_V + D_{VB} \nabla X_B),
\end{equation}
\begin{equation}
\frac{\partial X_B}{\partial t} = \nabla \cdot (D_{BV} \nabla X_V + D_{BB}^V \nabla X_B).
\end{equation}
\end{subequations}
Since the solid surfaces are very efficient vacancy sources/sinks \cite{Yu:2009a,Yu:2009b}, we impose the thermal-equilibrium vacancy mole fraction at the solid surfaces as the Dirichlet boundary condition for solving Eq.~\eqref{KE-Rig-1}. In this case, the smoothed boundary formulation of Eq.~\eqref{KE-Rig-1} is given by
\begin{subequations}
\begin{equation}
\psi^2 \frac{\partial X_V}{\partial t} = \psi \nabla \cdot [\psi (D_{VV} \nabla X_V+D_{VB} \nabla X_B)]-K,
\end{equation}
\begin{equation}
\psi^2 \frac{\partial X_B}{\partial t} = \psi \nabla \cdot [\psi (D_{BV} \nabla X_V+D_{BB}^V \nabla X_B)]+\frac{X_B}{1-X_V^{eq}}K,
\end{equation}
\end{subequations}
where $K = D_{VV} [\nabla \psi \cdot \nabla (\psi X_V)-|\nabla \psi|^2 X_V^{eq}]$. Since the vacancy generation and elimination in this scenario only occurs on the solid surfaces, no internal volume change needs to be considered in the bulk. Therefore, instead of using a plastic deformation model as in the previous case, we adopt a typical Cahn-Hilliard type dynamics to track the shape change:
\begin{equation}
\frac{\partial \psi}{\partial t} = M\nabla^2 \mu+\frac{\nabla \psi}{|\nabla \psi |}\cdot \frac{\vec{J}_V}{1-X_{V}^{eq}},
\end{equation}
where $\vec{J}_V = D_{VV}\nabla X_V + D_{VB}\nabla X_B$ is the vacancy flux, and the last term represents the normal velocity of the solid surfaces due to vacancy injection into or ejection from the solid.

\begin{figure}[htb]
\begin{center}
\includegraphics[width=1\textwidth]{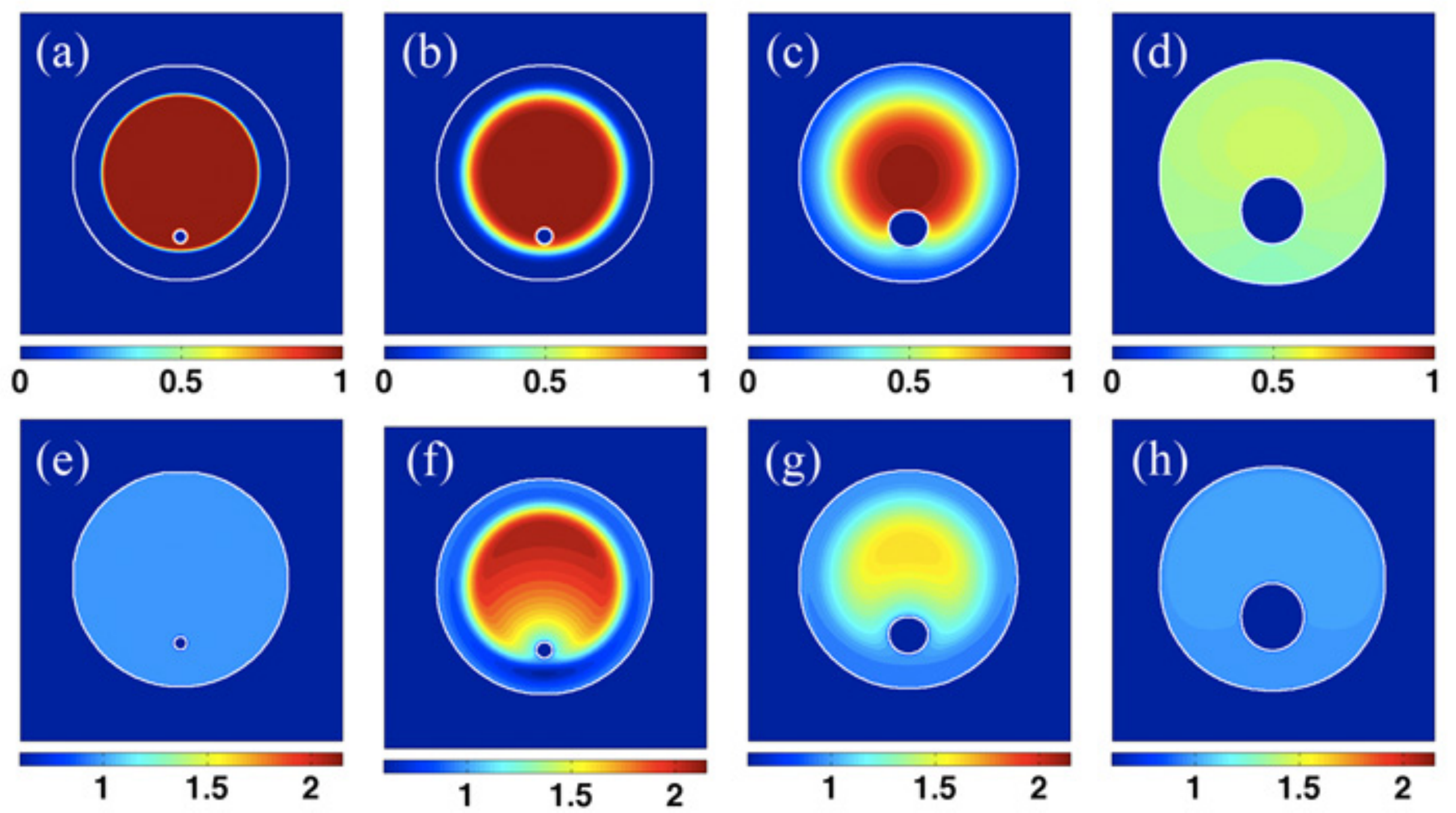}
\caption{Top row: snapshots of the fast diffuser mole fraction taken at 4 different times (a)$-$(d) from initial to final stages. Bottom row:  snapshots of vacancies mole fraction normalized to the equilibrium value taken at 4 different times, (e)$-$(h) corresponding to (a)$-$(d). The white solid contour lines indicate the locations of the rod and void surfaces.}\label{Hallow-1}
\end{center}
\end{figure}

An example of the results obtained using this approach is the growth of a void in a rod \cite{Fan:2006,Fan:2007,Yu:2009b}.  The above equations are solved using a central difference scheme in space and an implicit time stepping scheme. The vacancy mole fraction is fixed at the void and cylinder surfaces. The fast diffuser initially occupies the center region, while the slow diffuser occupies the outer region. A void is initially placed off-center in the fast diffuser region. Figure \ref{Hallow-1} shows snapshots of the fast diffuser mole fraction profile and the vacancy mole fraction profile (normalized to its equilibrium value). As the fast diffuser diffuses outward, vacancies diffuse inward from the rod surface to the void surface, causing vacancy concentration enhancement and depletion in the center and outer regions, respectively. To maintain the equilibrium vacancy mole fraction at the rod and void surfaces, vacancies are injected and ejected at those surfaces. As a result, the rod radius increases, and the void grows. Such dynamics was examined using a sharp interface approach \cite{Yu:2009b}, but this new method provides the flexibility in geometry to examine cases where a void initially forms off-center. 

\subsection{Thermal Stress}

Since an SOFC operates at temperatures near 500$\sim$1000$^\circ$C, the thermal stress is important for analyzing mechanical failure. Here, we expand the generalized mechanical equilibrium equation, Eq.~\eqref{SBM-ME-2}, for a linear, elastic and isotropic solid. In this case, the elastic constant tensor is expressed by
\begin{subequations}
\begin{equation}
\lambda_{11} = C_{1111} = C_{2222} = C_{3333},
\end{equation}
\begin{equation}
\lambda_{12} = C_{1122} = C_{2211} = C_{2233} = C_{3322} = C_{3311} = C_{1133},
\end{equation}
\begin{equation}
\begin{split}
\lambda_{44} =&  C_{1212} = C_{1221} = C_{2112} = C_{2121} = C_{2323} = C_{2332} \\= &C_{3223} = C_{3232} = C_{1313} = C_{1331} = C_{3113} = C_{3131}.
\end{split}
\end{equation}
\end{subequations}
The remainder of the elastic constant components vanish. We use the coordinate notation to replace the indices $i = 1$, $2$ and $3$ by $x$, $y$ and $z$, respectively. The infinitesimal displacements along the $x$, $y$ and $z$ directions are then replaced by $u$, $v$ and $w$, respectively. Thus, Eq.~\eqref{SBM-ME-2} in the three Cartesian directions is rewritten as
\begin{subequations} \label{ME-ISO-1}
\begin{equation} \label{ME-I1-3}
\begin{split}
\frac{\partial}{\partial x}\bigg[ \psi \lambda_{11} \bigg(\frac{\partial u}{\partial x} \bigg) \bigg] +
\frac{\partial}{\partial x}\bigg[ \psi \lambda_{12} \bigg( \frac{\partial v}{\partial y} \bigg) \bigg] +
\frac{\partial}{\partial x}\bigg[ \psi \lambda_{12} \bigg( \frac{\partial w}{\partial z} \bigg) \bigg] & + \\
\frac{\partial}{\partial y}\bigg[ \psi \lambda_{44} \bigg( \frac{\partial u}{\partial y} + \frac{\partial v}{\partial x} \bigg) \bigg] + 
\frac{\partial}{\partial z}\bigg[ \psi \lambda_{44} \bigg( \frac{\partial u}{\partial z} + \frac{\partial w}{\partial x} \bigg) \bigg] & + \\ |\nabla \psi | N_x =   \frac{\partial}{\partial x} [ \psi \rho (\lambda_{11}+2 \lambda_{12}) ] &,
\end{split}
\end{equation}
\begin{equation} \label{ME-I2-3}
\begin{split}
\frac{\partial}{\partial x}\bigg[ \psi \lambda_{44} \bigg( \frac{\partial u}{\partial y} + \frac{\partial v}{\partial x} \bigg) \bigg] & + \\ \frac{\partial}{\partial y}\bigg[ \psi \lambda_{12} \bigg( \frac{\partial u}{\partial x} \bigg) \bigg] +
\frac{\partial}{\partial y}\bigg[ \psi \lambda_{11} \bigg( \frac{\partial v}{\partial y} \bigg) \bigg] +
\frac{\partial}{\partial y}\bigg[ \psi \lambda_{12} \bigg( \frac{\partial w}{\partial z} \bigg) \bigg] & + \\
\frac{\partial}{\partial z}\bigg[ \psi \lambda_{44} \bigg( \frac{\partial v}{\partial z} + \frac{\partial w}{\partial y} \bigg) \bigg]  & + \\ |\nabla \psi | N_y= \frac{\partial}{\partial y} [  \psi \rho (\lambda_{11}+2 \lambda_{12}) ] &,
\end{split}
\end{equation}
\begin{equation} \label{ME-I3-3}
\begin{split}
\frac{\partial}{\partial x}\bigg[ \psi \lambda_{44} \bigg( \frac{\partial u}{\partial z} + \frac{\partial w}{\partial x} \bigg) \bigg] +  \frac{\partial}{\partial y}\bigg[ \psi \lambda_{44} \bigg( \frac{\partial v}{\partial z} + \frac{\partial w}{\partial y} \bigg) \bigg] &+\\
\frac{\partial}{\partial z}\bigg[ \psi \lambda_{12} \bigg( \frac{\partial u}{\partial x} \bigg) \bigg] +
\frac{\partial}{\partial z}\bigg[ \psi \lambda_{12} \bigg( \frac{\partial v}{\partial y} \bigg) \bigg] +
\frac{\partial}{\partial z}\bigg[ \psi \lambda_{11} \bigg( \frac{\partial w}{\partial z} \bigg) \bigg] & + \\ |\nabla \psi |N_z= \frac{\partial}{\partial z} [ \psi \rho (\lambda_{11}+2 \lambda_{12}) ] &,
\end{split}
\end{equation}
\end{subequations}
for the $x$, $y$ and $z$ directions, respectively. To numerically solve this equation, we reorganize the terms in Eqs.~\eqref{ME-I1-3}--\eqref{ME-I3-3} to form
\begin{equation} \label{LDOP-1}
\mathcal{L}_1u = h_1,~\mathcal{L}_2v = h_2,~\text{and}~\mathcal{L}_3w = h_3,
\end{equation}
where $\mathcal{L}_1$, $\mathcal{L}_2$ and $\mathcal{L}_2$ are the linear differential operators associated with $u$, $v$ and $w$, respectively, in Eq.~\eqref{ME-ISO-1}. The right-hand sides, $h_1$, $h_2$ and $h_3$, are the remaining terms collected in Eq.~\eqref{ME-ISO-1}. The linear differential operators are discretized in the second-order central differencing scheme in space. We employ an ADI solver for the linear differential operators and iterate Eq.~\eqref{LDOP-1} until $u$, $v$ and $w$ all converge to their equilibrium values. 

We select the material parameters as follows: $\alpha^{YSZ} \Delta T = 1\%$ and $\alpha^{LSC}\Delta T= 2\%$. The elastic constants are choosen arbitrarily as $\lambda_{11}^{YSZ}=20\times10^7$, $\lambda_{22}^{YSZ}=10\times10^7$, and $\lambda_{44}^{YSZ} = 5\times10^7$ (dimensionless) such that the solid is isotropic in mechanical behavior, $(\lambda_{11}-\lambda_{12})/(2\lambda_{44})=1$. The LSC (cathode) phase is softer than the YSZ (electrolyte) phase, and its elastic constant is assumed to be $0.75\lambda_{ij}^{YSZ}$. Again, we use domain parameters to indicate the YSZ phase ($\psi_{YSZ} = 1$ inside YSZ and $\psi_{YSZ} = 0$ outside YSZ) and LSC phase ($\psi_{LSC}=1$ inside LSC and $\psi_{LSC}=0$ outside LSC). The entire solid phase is then indicated by the sum of the two phases, $\psi = \psi_{YSZ}+\psi_{LSC} = 1$. The body force term and elastic constant tensor in Eq.~\eqref{ME-ISO-1} are replaced by an interpolated, spatially dependent thermal expansion, $\psi \rho = 0.01\psi_{YSZ} + 0.02\psi_{LSC}$, and elastic constant tensor, $\lambda_{ij} = \lambda_{ij}^{YSZ}\psi_{YSZ}+\lambda_{ij}^{LSC}\psi_{LSC}$. The solid surface is assumed to be traction-free, $N_i=0$. 

In this simulation, we select a computational box containing 160, 160, and 149 grid points along the $x$, $y$, and $z$ directions, respectively. The grid spacing is $\Delta x = 0.04$. We assume a rigid computational box with frictionless boundaries on the six sides, which means that $u=0$, $v$ and $w$ are free on the two $y$-$z$ planes, $v=0$,  $u$ and $w$ are free on the two $x$-$z$ planes, and $w=0$, $u$ and $v$ are free on the two $x$-$y$ planes of the computational box boundaries.

\begin{figure}[htb]
\begin{center}
\includegraphics[width=1\textwidth]{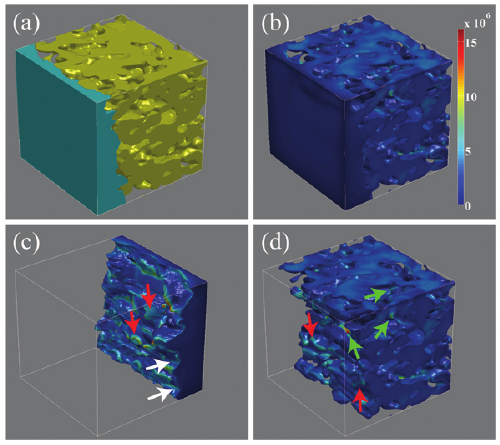}
\caption{(a) Solid phase containing cathode (yellow) and electrolyte (cyan) in SOFC. (b) Von-Mises stress in the entire solid phase due to thermal expansion. (c) Von-Mises stress in the electrolyte phase. (d) Von-Mises stress in the cathode phase.}\label{TherStress}
\end{center}
\end{figure}

Figure \ref{TherStress}(a) illustrates our experimentally obtained microstructure containing the cathode (LSC) and electrolyte (YSZ) phases. The yellow color indicates the cathode phase, and the cyan color indicates the electrolyte phase. Shown in Fig.~\ref{TherStress}(b) are the calculated Von-Mises stresses resulting from the thermal expansion. Due to the porosity, an overall stress enhancement occurs in the cathode phase, as can be observed from the overall light blue-green color. Figure \ref{TherStress}(c) shows the Von-Mises stress on the YSZ surface after rotating the volume 180$^\circ$ around the $z$-axis. Figure \ref{TherStress}(d) shows the Von-Mises stress in the LSC phase.

Three types of stress enhancements can be noticed from the simulation result. At the cathode-electrolyte contacting surfaces, stress is enhanced due to the mismatch of thermal expansion and elastic constants between the two materials (see the red arrows in Figs.~\ref{TherStress}(c) and (d)). The second is the concentrated stress observed at the grooves on the electrolyte surface (not contacting the cathode) as shown in the white arrows in Fig.~\ref{TherStress}(c). The third type is the stress concentration effect at the bottlenecks in the cathode phase as shown in the green arrows in Fig.~\ref{TherStress}(d). The simulation results demonstrate that the smoothed boundary method can properly capture the linear elasticity behavior and the geometric effects based on a diffuse-interface defined geometry.

\subsection{Phase Transformations in the Presence of a Foreign Surface} \label{Contact angle AC}

\begin{figure}[htbp]
\begin{center}
\includegraphics[width=1\textwidth]{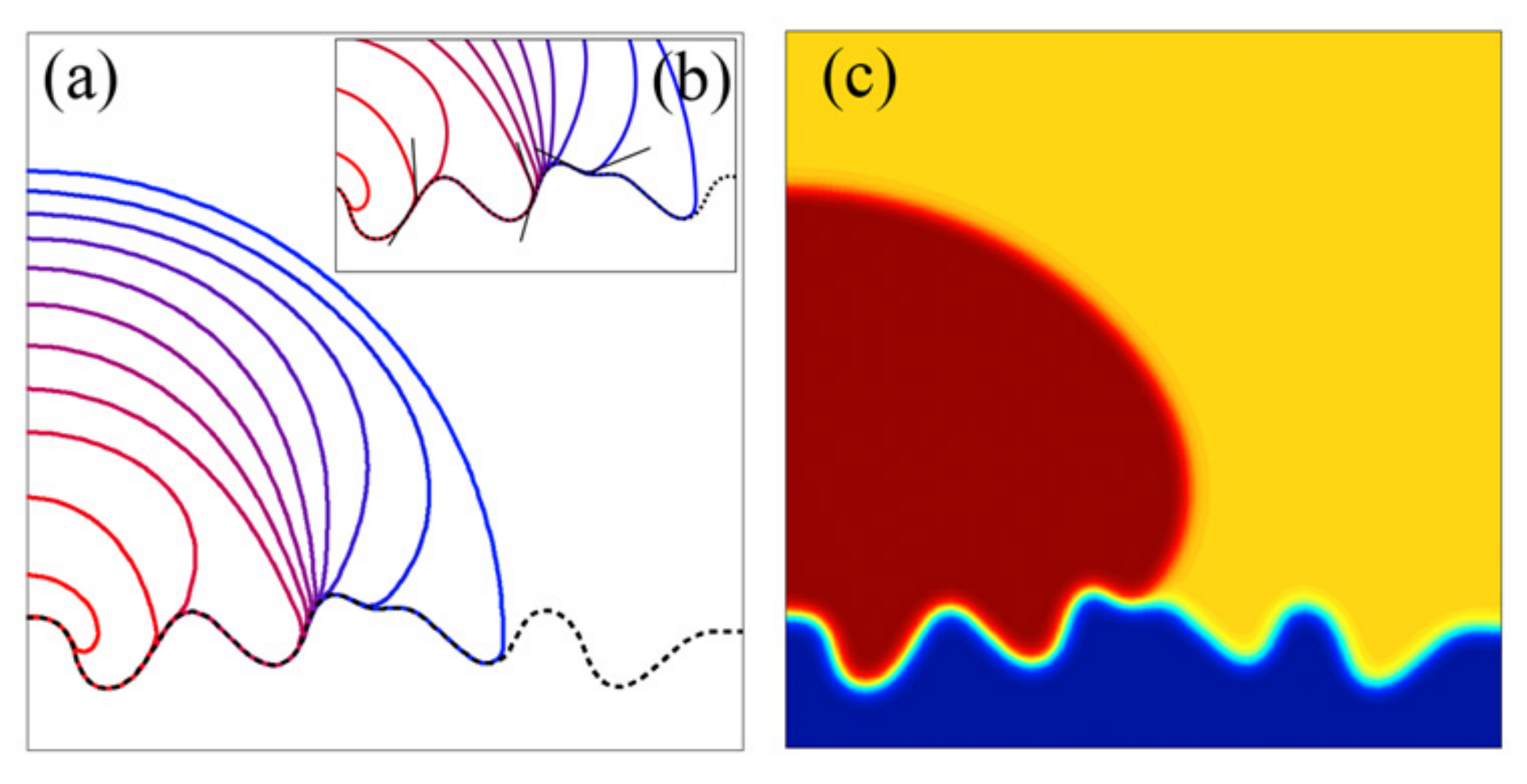}
\caption{(a) The dynamics of evaporation of a droplet on a rough surface (dashed line) governed by Allen-Cahn dynamics. The contact angle between the droplet and the surface is imposed at 135 degrees.  Solid curves with various colors represent the profile of the droplet at different times. The outermost blue line represents the initial state, and the innermost represents the final state (taken before complete evaporation in the simulation); each line is plotted at every time interval of 270 in dimensionless time.  The velocity of the three-phase boundary is greatly affected by the surface profile. (b) A magnified view of the three-phase boundary to show the contact angle is accurately set (the angle made by the thin black lines is 135 degrees. (c) The order and domain parameters are shown to illustrate the diffuse nature of the interface and boundary (taken at $t = 270$).}\label{droplet}
\end{center}
\end{figure}

The Allen-Cahn equation describes the dynamics of a nonconserved order parameter, which can be taken as a model for the ordering of magnetic moments \cite{Chen:2002} and diffusionless phase transformations that involve only crystalline order change \cite{Chen:2002}.  It can also be used as a model for evaporation-condensation dynamics \cite{Chen:2002,Emmerich:2003}.  Here, we use the Allen-Cahn equation to examine the evaporation of a droplet on a rough surface. The domain parameter is given a ripple-like feature as shown in Fig.\ \ref{droplet}, and pre-smoothed using Allen-Cahn dynamics, Eq.~\eqref{S-1}. The droplet phase is placed on top of the boundary, and its shape is evolved by the smoothed boundary formulation of the Allen-Cahn equation, Eq.\ \eqref{AC-1}.  The simulation is performed in two dimensions using parameters $\Delta x = 1$, $M=1$ and $\epsilon=1$ with a domain size of $L_x=100$ and $L_y=100$.

The evolution of the droplet surface as it evaporates is illustrated in Fig.\ \ref{droplet}(a) as a contour ($\phi=0.5$) plotted at equal intervals of 270 in dimensionless time.  The blue to red colors indicate the initial to final stages. As it evolves, it is clear that the contact angle is maintained, as shown in Fig.\ \ref{droplet}(b). The dynamics of the motion of the three-phase boundary is interesting in that the velocity changes depending on the angle of the surface (with respect to the horizontal axis), which can be inferred from the change in the density of the contours.  Since the interfacial energy is assumed to be constant, the droplet would prefer to have a circular cap shape.  However, the contact angle imposes another constraint at the three-phase boundary.  When the orientation of the surface is such that both of these conditions are nearly met, the motion of the three-phase boundary is slow while the droplet continues to evaporate.  When the orientation becomes such that the shape of the droplet near the three-phase boundary must be deformed (compared to the circular cap), the three-phase boundary moves very quickly.  This leads to an unsteady motion of the three-phase boundary. On the other hand, at the top of the droplet far from the substrate, the curvature is barely affected by the angle of the substrate surface; thus, the phase boundary there moves at a speed inversely proportional to the radius.

\subsection{Motion of a Droplet due to Unbalanced Surface Tensions} \label{Contact angle CH}

As another application, we have modeled a self-propelling droplet.  Here, two different contact-angle boundary conditions are imposed on the right and left sides of the droplet placed on a flat surface.  The smoothed boundary formulation of the Cahn-Hilliard equation, Eq.~\eqref{SBM-CH-3}, is used with $J_{n} = 0$ in this simulation.  The parameters used are $\Delta x = 1$, $M=1$, and $\epsilon=1$, and the domain sizes are $L_x=240$ and $L_y=60$. The contact angle on the right side of the droplet is set to 45 degrees and that on the left side to 60 degrees by imposing position dependent boundary conditions. Note that this setup is equivalent to the situation in which the wall-environment, droplet-wall and droplet-environment surface energies satisfy the conditions of Young's equation as
\begin{subequations}
\begin{equation}
\gamma_{we}-\gamma_{wd} = \gamma_{de} \cos{60^{\circ}} ~~\text{for the left side},
\end{equation}
\begin{equation}
\gamma_{we}-\gamma_{wd} = \gamma_{de} \cos{45^{\circ}}~~\text{for the right side},
\end{equation}
\end{subequations}
where $\gamma_{we}$, $\gamma_{wd}$ and $\gamma_{de}$ are the surface energies of the wall-environment, droplet-wall and droplet-environment surfaces, respectively, Therefore, this model can be used to simulate a case where the surface energies are spatially and/or temporally dependent on other fields, such as surface temperature or surface composition, as in Ref \cite{Tersoff:2009}.

\begin{figure}[htbp]
\begin{center}
\includegraphics[width=1\textwidth]{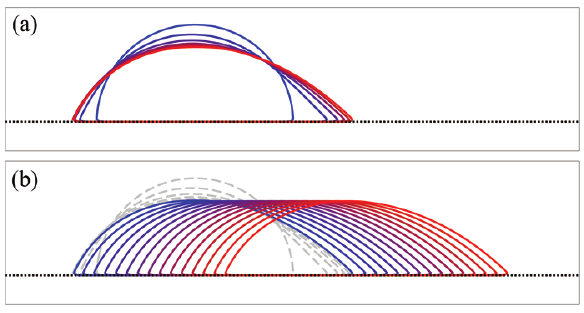}
\caption{A self-propelling droplet driven by unbalanced surface tensions. The evolution is modeled by the Chan-Hilliard equation with two different contact-angle BCs on each side of the droplet. (a) The droplet shape changes during the relaxation period. The color contours are plotted at time intervals of $2\times10^{4}$ in dimensionless times. (b) The droplet motion along the substrate surface. The color contours are plotted at time intervals of $1\times10^{5}$ in dimensionless time. The droplet moves at constant speed in the steady state.}\label{self-propel-droplet}
\end{center}
\end{figure}

The evolution of the droplet surface is illustrated in Fig.~\ref{self-propel-droplet}. The droplet initially has the shape of a hemisphere, with a 90-degree contact angle with the wall surface. The early evolution is marked by the evolution of the droplet shape as it relaxes to satisfy the contact-angle boundary condition, as seen in Fig.~\ref{self-propel-droplet}(a). Then the droplet begins to accelerate. Once the contact angle reaches the prescribed value, it is maintained as the droplet moves toward the right (see Fig.~\ref{self-propel-droplet}(b)). In the steady state, the droplet moves at constant speed without other effects present. Such motions of droplets have been observed and explained as a result of an unbalanced surface tension between the head portion (with a dry surface) and tail portion (with a wet surface) due to the resulting spatially varying composition and composition-dependent surface energy \cite{Tersoff:2009}.  

\begin{figure}[htbp]
\begin{center}
\includegraphics[width=1\textwidth]{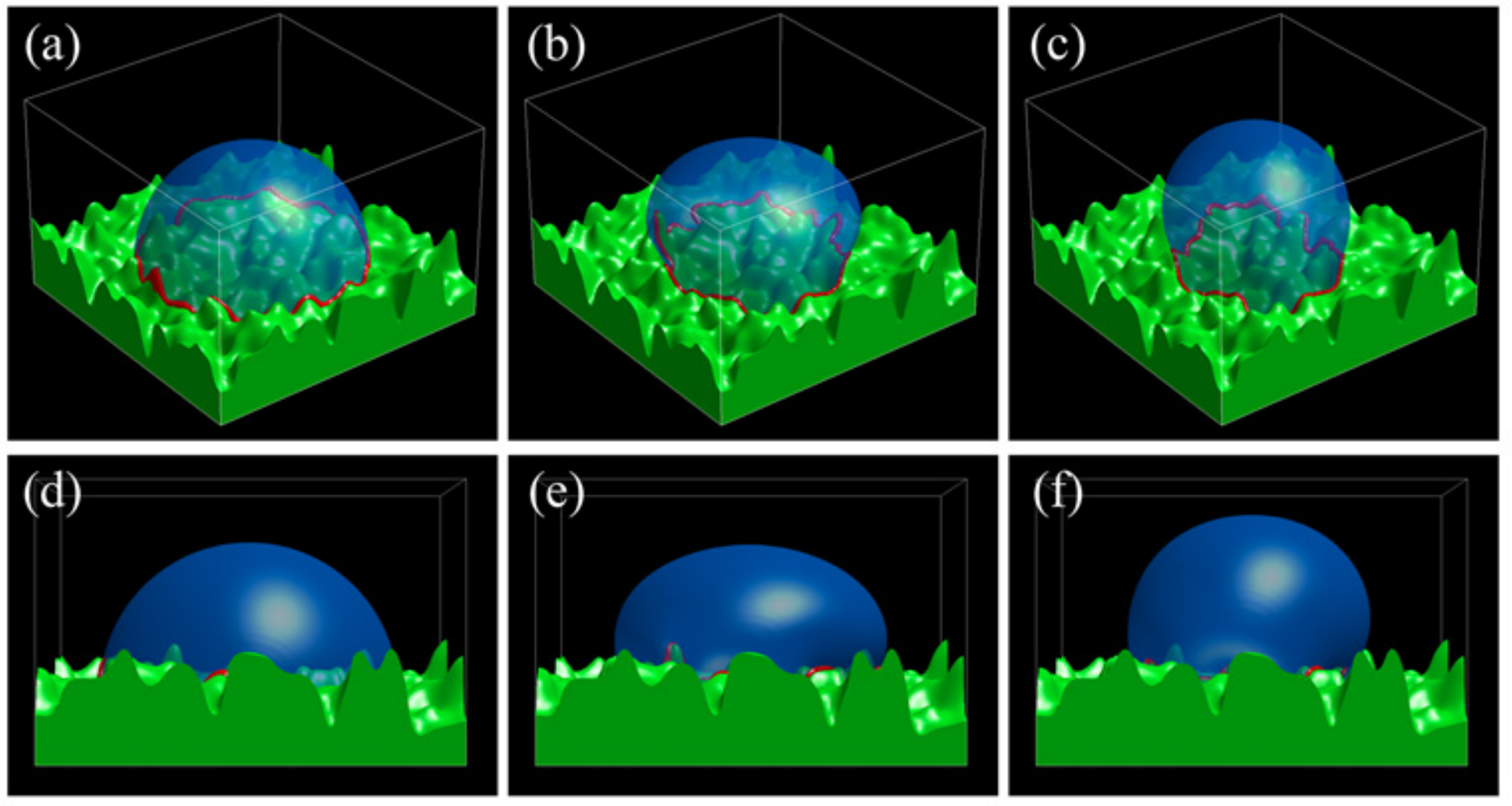}
\caption{A droplet relaxing toward the equilibrium shape. The evolution is modeled by the Chan-Hilliard equation with a contact angle of 135$^\circ$ to the irregular substrate surface: (a) initial ($t=0$), (b) intermediate ($t=3\times10^{3}$), and (c) equilibrium state ($t=2.35\times10^{4}$). The three-phase boundaries are illustrated by the red color. The side views of the droplet are shown in (d), (e) and (f) corresponding to (a), (b) and (c), respectively. } \label{relaxingdroplet}
\end{center}
\end{figure}

Figure \ref{relaxingdroplet} shows the relaxation of an initially hemispherical droplet on an irregular substrate surface in a 3D simulation. The contact-angle boundary condition imposed at the three-phase boundary is 135 degrees. The computational box sizes are $L_x=L_y=120$ and $L_z= 80$. As can be seen, the droplet changes its shape to satisfy the imposed contact angle, and the droplet evolves to a shape for which the total surface energy is minimized. The behavior favoring dewetting imposed by the contact angle ($\theta > 90^{\circ}$) is properly reflected in the lifting of the droplet, as shown in Figs.~\ref{relaxingdroplet}(a)--(c) and (d)--(f). During this relaxation process, the three-phase boundary shrinks toward the center as the droplet-wall contacting area decreases, as shown in Fig.~\ref{relaxingdroplet}(a)--(c).

\section{Discussions and Conclusions}

In this paper, we have demonstrated a generalized formulation of the smoothed boundary method. This method can properly impose Neumann and/or Dirichlet boundary conditions on a diffuse interface for solving partial differential equations within the region where the domain parameter $\psi$ uniformly equals 1. The derivation of the method, as well as its implementation, is straightforward. It can numerically solve differential equations without complicated and time-consuming meshing of the domain of interest since the domain boundary is specified by a spatially varying function. Instead, any gird system, including a regular Cartesian grid system, can be used with this method.

This smoothed boundary approach is flexible in coupling multiple differential equations. We have demonstrated how this method can couple bulk diffusion and surface diffusion into one single equation while the two equations serve as the boundary condition for each other in Section \ref{SurfDiffFormulation}. In principle, this method can couple multiple differential equations in different regions that are defined by different domain parameters. For example, the physics within a domain defined by $\psi_i = 1$ is governed by a differential equation $H_i$. The overall phenomenon will be then represented by $H = \sum_{i} \psi_i H_i$, where the subscript `$i$' denotes the $i$-$th$ domain, and $\sum_{i} \psi_i = 1$ represents the entire computational box. When sharing the diffuse interfaces between domains, the physical quantities can connect to one another as boundary conditions for each equation in each domain. Therefore, this method could be used to simulate coupled multi-physics and/or multiple-domain problems, such as fluid-solid interaction phenomena or diffusion in hetero-polycrystalline solids.
 
We have also demonstrated the capability of applying the smoothed boundary method to moving boundary problems in Section \ref{Kirkendall effect deformation}. When the locations of domain boundaries are updated by a phase-field type dynamics such that the domain parameter remains uniformly at 1 and 0 on each side of the interface, the smoothed boundary method can be conveniently employed to solve differential equations with moving boundaries. 

In addition to Neumann and Dirichlet boundary conditions, we have also shown the capability of the smoothed boundary method for specifying contact angles between the phase boundaries and domain boundaries (Sections \ref{Contact angle AC} and \ref{Contact angle CH}). This type of boundary condition is difficult to impose using conventional sharp interface models.

Although the smoothed boundary method has many advantages, as shown in Section \ref{DiffEqn}, the nature of the diffuse interface inevitably introduces an error proportional to the interfacial thickness since we smear an originally zero-thickness boundary into a finite thickness interface. Another error results from the resolution of the rapid transition of the domain parameter across the interfacial region. When numerically solving the smoothed-boundary formulated equations, properly capturing the gradient of the domain parameter across the interface becomes very important. From our experience, at least 4$\sim$6 grid points are necessary to resolve the diffuse interfaces such that the errors are controlled. Moreover, when solving time dependent equations, one singularity occurs because of the terms of $1/\psi$ and $1/\psi^2$ for imposing Neumann and Dirichlet boundary conditions, respectively. In practice, cutoffs at small $\psi$ values are necessary to avoid numerical instabilities. These cutoff values can be smaller as the diffuse interface is better resolved, i.e., by using more grid points across the interface. However, only a small number of grid points will be used across the interface for computational efficiency. In our simulations, when 4$\sim$6 grid spacings are used for the interfacial regions, the cutoff values are around $1\times10^{-6}\sim1\times10^{-8}$ for the Neumann boundary condition and $1\times10^{-2}\sim1\times10^{-4}$ for the Dirichlet boundary condition to maintain numerical stability while keeping the errors reasonably small. On the other hand, when solving time independent equations such as the mechanical equilibrium equation and the steady state diffusion equation, there are no singular terms in the equations. The cutoff value is simply used to avoid the singularity of the matrix solver. In this case, the cutoff value can be as small as the order of numerical precision, such as $1\times10^{-16}$. All of these numerical instability and error behaviors require more systematic and theoretical studies; thus, the interfacial thickness and resolution should be optimized for future works.

 Based on the general nature of the derivation, the smoothed boundary method is applicable to generalized boundary conditions (including time-dependent boundary values that are important for simulating evolution of many physical systems). Since the domain boundaries are not specifically defined in the smoothed boundary method, this method can be applied to almost any geometry as long as it can be defined by the domain parameter. This is very powerful and convenient for solving differential equations in complex geometries that are often difficult and time-consuming to mesh. As three-dimensional image-based calculations are more prevailing in scientific and engineering research fields \cite{Thornton:2008} in which voxelated data from serial scanning or sectioning are often utilized and are difficult to render as meshes, the smoothed boundary method is expected to be widely employed to simulate and study physics in complex geometries defined by 2D pixelated and 3D voxelated data with a simply process of smoothing the domain boundaries.

\textbf{Acknowledgements}: HCY and KT thank the National Science Foundation for financial support under Grant Nos. 0511232, 0502737, and 0854905.  HYC and KT thank the National Science Foundation for financial support on Grant Nos. 0542619 and 0907030.  The authors thank John Lowengrub, Axel Voigt, Xiaofan Li, Anton Van der Ven and James Warren for valuable discussions and comments.  The authors also thank Scott Barnett and Stu Adler for providing the experimental 3D microstructures used in the demonstration.


\newpage

\end{document}